\documentclass{emulateapj}

\usepackage{natbib}

\shorttitle{The Environments of Local SNe Ia}
\shortauthors{Cooper, Newman, \& Yan}

\begin{document}



\title{The Large--Scale Environments of Type I\lowercase{a}
  Supernovae: Evidence for a Metallicity Bias in the Rate or
  Luminosity of Prompt I\lowercase{a} Events}


\author{
Michael C.\ Cooper\altaffilmark{1,2},
Jeffrey A.\ Newman\altaffilmark{3}, 
Renbin Yan\altaffilmark{4}
}

\altaffiltext{1}{Steward Observatory, University of Arizona, 
933 N.\ Cherry Avenue, Tucson, AZ 85721 USA; 
cooper@as.arizona.edu}

\altaffiltext{2}{Spitzer Fellow}

\altaffiltext{3}{Department of Physics and Astronomy, University of
  Pittsburgh, 401--C Allen Hall, 3941 O'Hara Street, Pittsburgh, PA 15260
  USA; janewman@pitt.edu}

\altaffiltext{4}{Department of Astronomy and Astrophysics, University
  of Toronto, 50 St.\ George Street, Toronto, ON M5S 3H4, Canada;
  yan@astro.utoronto.ca}

\begin{abstract}

  Using data drawn from the Sloan Digital Sky Survey (SDSS) and the
  SDSS--II Supernova Survey, we study the local environments of
  confirmed type Ia supernovae (SNe Ia) in the nearby Universe. At
  $0.05 < z < 0.15$, we find that SN Ia events in blue, star--forming
  galaxies occur preferentially in regions of lower galaxy density
  relative to galaxies of like stellar mass and star--formation rate,
  while SNe Ia in nearby red galaxies show no significant environment
  dependence within the measurement uncertainties. Even though our
  samples of SNe in red hosts are relatively small in number, tests on
  simulated galaxy samples suggest that the observed distribution of
  environments for red SN Ia hosts is in poor agreement with a cluster
  type Ia rate strongly elevated relative to the field rate. Finally,
  after considering the impact of galaxy morphology, stellar age,
  stellar metallicity, and other relevant galaxy properties, we
  conclude that the observed correlation between the SN Ia rate and
  environment in the star--forming galaxy population is likely driven
  by a gas--phase metallicity effect, such that prompt type Ia
  supernovae occur more often or are more luminous in metal--poor
  systems.

\end{abstract}


\section{Introduction}
\label{sec_intro}

Type Ia supernovae (SNe Ia) are thought to be distinct from other
types of supernovae (that is, resulting from a different progenitor
population), as they are found in galaxies spanning a broad range of
properties. While, type II, Ib, and Ic SNe are only found in
star--forming galaxies, indicating that they are the evolutionary
product of massive stars, SNe Ia are found in both star--forming and
quiescent systems, suggesting that they are somehow connected to the
evolution of less--massive stars \citep[e.g.,][]{oemler79, vdb90,
  dv94, cappellaro99}. In particular, it is widely accepted that the
progenitors of type Ia supernovae are carbon--oxygen white dwarfs
(WDs), which have accreted mass up to the Chandrasekhar limit
\citep{chandra31}, perhaps via deposition from a binary companion
\citep[][]{whelan73, han04}. Lending support to this picture,
theoretical models of exploding WDs \citep[e.g.,][]{kasen05, kasen07}
are able to reproduce the properties of SN Ia spectra, including the
lack of hydrogen features and the presence of strong silicon
absorption, which define the type Ia classification.

Recent observations of the type Ia SN rate in local and
intermediate--redshift galaxies have cast some doubt on --- or at
minimum confused --- the theoretical paradigm just discussed. If SNe
Ia are simply the product of old stellar populations (i.e., an
evolutionary outcome of low--mass stars), then the SN Ia rate should
depend strongly on stellar mass, while being independent of the level
of on--going star--formation activity (at fixed stellar mass). While
the type Ia rate is found to depend on stellar mass, it is also found
to be greater per unit stellar mass in galaxies with higher specific
star--formation rates \citep{mannucci05, sullivan06b}. These current
observations of the supernova Ia rate support a revised theoretical
model \citep[e.g.,][]{scannapieco05, mannucci06, pritchet08}, perhaps
one employing a two--component progenitor distribution with a
``prompt'' component correlated with star--formation activity and a
``delayed'' component correlated with stellar mass (i.e., the
underlying older stellar population). Still, despite uncertainty in
the nature of the two observed components of the type Ia rate as well
as difficulties in directly observing the progenitors of type Ia
events, current observations remain consistent with a progenitor
population comprised entirely of carbon--oxygen WDs, allowing for an
increasingly broad distribution of delay times
\citep[e.g.,][]{greggio08}.

Studying the environments of supernova --- i.e., their place in the
hierarchy of large--scale structure --- may prove useful in shedding
light on the nature of the two components of the type Ia rate or on
the nature of the underlying progenitor population or potentially even
revealing dependencies of the type Ia rate or luminosity on galaxy
properties that are correlated with environment. Type Ia supernovae
are often used as cosmological probes, allowing distances to be
measured out to intermediate redshifts and providing constraints on
cosmological parameters such as $H_0$, $\Omega_{M}$, and
$\Omega_{\Lambda}$ \citep{riess98, perlmutter99}. Understanding any
relationship between host galaxy properties and type Ia luminosity is
critical for minimizing systematic effects that might bias studies of
distant type Ia supernovae and limit our abilities to constrain
cosmological models. Along these lines, environment can be utilized as
a proxy for other galaxy properties that are otherwise difficult to
constrain observationally (e.g., stellar metallicity or age). For
instance, a galaxy of the same stellar mass as another, but located in
a higher--density environment, will generally have formed earlier in
the history of the Universe and will contain older stellar populations
as a result \citep[e.g.,][]{schaye03, dave06}.

Understanding if the type Ia rate varies with the local galaxy
environment may provide new insights into the true nature of the
progenitor population, while also shedding light on many aspects of
galaxy formation and evolution. For instance, using environment as a
proxy for formation time could be helpful in interpreting the role of
an evolving initial stellar mass function \citep[e.g.,][]{dave08,
  vandokkum08} in establishing the potential progenitor population.

More directly, constraining the correlation between the SN rate (of
both type Ia and II events) and galaxy environment is also critical
for understanding various details of the chemical enrichment and
star--formation histories of galaxies. For example, supernovae govern
the production of metals \citep{woosley95, sato07}, with types Ia and
II dominating iron and oxygen production, respectively. Due to their
large gravitational potential wells, galaxy clusters are potentially
the only systems to have retained all of the metals produced by
stars. For this reason, studies of the metal content (e.g., Fe
abundance) in the intracluster medium (ICM) provide interesting
windows on the cosmic star--formation history
\citep[e.g.,][]{matteucci88, calura07}. Given the role of supernovae
in dispersing metals, a critical part of this picture is to understand
how the SN rate may vary with environment in addition to time
\citep{suresh08}.

Furthermore, energy injected into the interstellar medium (ISM) by
supernovae could play an important role in influencing the formation
of galactic disks \citep{cscannapieco08}. Variation in the supernova
rate with environment could thus be a factor (though likely very weak)
in the establishment of the morphology--density relation
\citep[e.g.,][]{davis76, dressler80}. Finally, feedback from
supernova--generated winds is thought to be an essential part of
shaping the mass--metallicity relation \citep{dekel86, cole91},
causing a downturn in the relative enrichment at low stellar masses by
ejecting metals into the intergalactic medium (IGM). Understanding the
correlation between supernova rates and environment will help reveal
whether supernova feedback is also responsible for driving the scatter
in this fundamental relation, causing the observed correlation between
metallicity and environment at fixed stellar mass and star--formation
rate \citep{cooper08b}.

In this paper, we utilize public data from the Sloan Digital Sky
Survey \citep[SDSS,][]{york00} and from the SDSS--II Supernova Survey
\citep{frieman08} to examine the environments of local type Ia SNe.
In \S 2, we discuss the data samples employed along with our
measurements of galaxy environments. Our main results regarding the
environments of type Ia SNe are presented in \S 3, with comparison to
related work, analysis of potential selection effects, and further
discussion in \S 4. Finally, in \S 5, we summarize our
conclusions. Throughout this paper, we assume a flat $\Lambda$CDM
cosmology with $\Omega_m = 0.3$, $\Omega_{\Lambda} = 0.7$, $w = -1$,
and a Hubble parameter of $H_0 = 100\ {\rm km}\ {\rm s}^{-1}\ {\rm
  Mpc}^{-1}$, unless otherwise noted.

\section{The Data Sets}
\label{sec_data}

\subsection{The Supernova Sample}

Over the past three years, the SDSS--II Supernova Survey
\citep{frieman08, sako08} has repeatedly scanned a $300$ square degree
region around the celestial equator in the southern Galactic
hemisphere in search of supernovae. The resulting data set, which
includes more than $800$ SNe, is exceptional in its size and
uniformity, significantly increasing the number of known supernovae at
$z \lesssim 0.3$ with well--sampled and well--calibrated light curves.
Only the Supernova Legacy Survey \citep[SNLS,][]{astier06}, which
primarily targets supernovae at higher redshifts $(z > 0.3)$, has
discovered a comparable number of type Ia events.

In this work, we utilize the $> \! 500$ confirmed SNe Ia from the
SDSS--II Supernova Survey.\footnote{Downloaded from
  http://sdssdp47.fnal.gov/sdsssn/sdsssn.html.} We limit the sample to
a redshift range of $0.05 < z < 0.15$ and to SNe for which a spectrum
was obtained, which yields a total sample size of 163 type Ia SNe
(SN--A sample). We exclude type II supernovae from our analysis due to
their limited numbers; there is only one type II supernova for every
$\sim \! 10$ type Ia event in the SDSS--II Supernovae Survey
catalog. While type II events are more numerous than SNe Ia
\citep[e.g.,][]{maoz08}, they are intrinsically fainter and thus
less--likely to be detected in a magnitude--limited survey.

A key aspect of the SDSS--II Supernova Survey is the general spatial
uniformity of the number of imaging epochs. That is, there is little
spatial dependence to the on--sky cadence of the survey \citep[see
Figure 1 of][and \citealt{sako08}]{frieman08}. This translates to a
survey that is equally sensitive to supernovae in all environments,
which therefore allows for direct comparison of the environments of SN
hosts to the environments of the global galaxy population.

\subsection{The Galaxy Sample}

With spectra and multi--band photometry for more than $500,000$
galaxies, the SDSS Data Release 6 \citep[DR6,][]{adelman08} enables the
local density of galaxies (which we refer to throughout this paper as
``environment'') at $z \lesssim 0.2$ to be measured over approximately
one quarter of the sky, including nearly the entire area surveyed by
the SDSS--II Supernova Survey. We select a sample of $553,188$
galaxies from the SDSS DR6, as contained in the New York University
Value--Added Galaxy Catalog \citep[NYU--VAGC,][]{blanton05b}. This
sample is limited to the redshift range $0.01 < z < 0.3$ and to SDSS
fiber plates for which the redshift success rate for targets in the
main spectroscopic survey is 80 per cent or greater.

For each galaxy, the SDSS/NYU--VAGC database provides precise
information about position on the plane of the sky ($< \! 0.1$
arcsecond rms per coordinate) and along the line of sight
\citep[$\Delta z \sim 0.0001$,][]{abazajian04}, enabling the local
environment to be accurately characterized (see \S
\ref{sec_data_environ}). In addition, rest--frame colors and absolute
magnitudes are computed using the KCORRECT $K$--correction code
(version v4\_1\_4) of \citet[][see also
\citealt{blanton03}]{blanton07}. The rest--frame quantities for the
SDSS sample are derived from the apparent $ugriz$ model magnitudes in
the SDSS DR6, where all SDSS magnitudes within this paper are
calibrated to the AB system \citep{oke83}. 

\begin{figure*}[t]
\centering
\plottwo{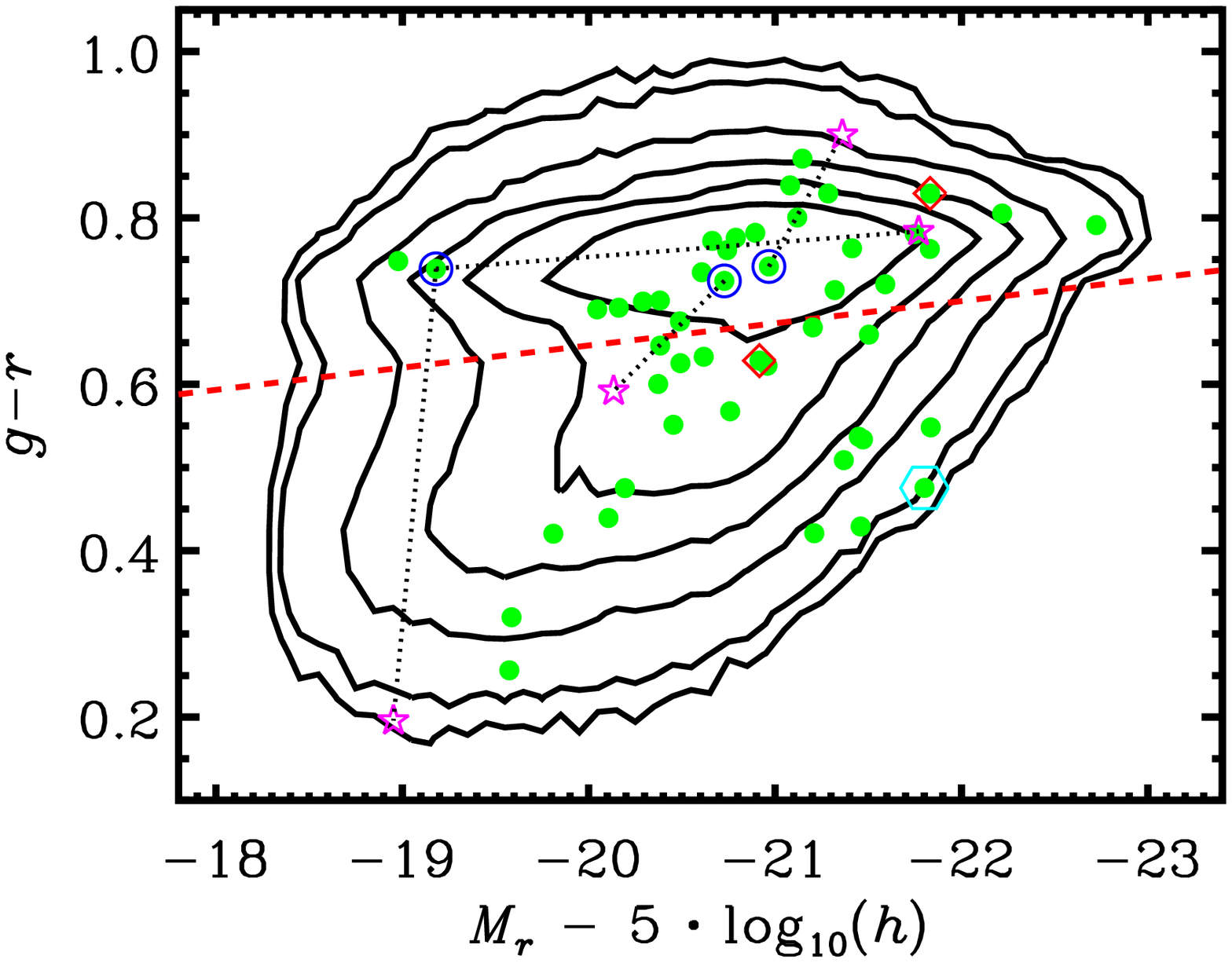}{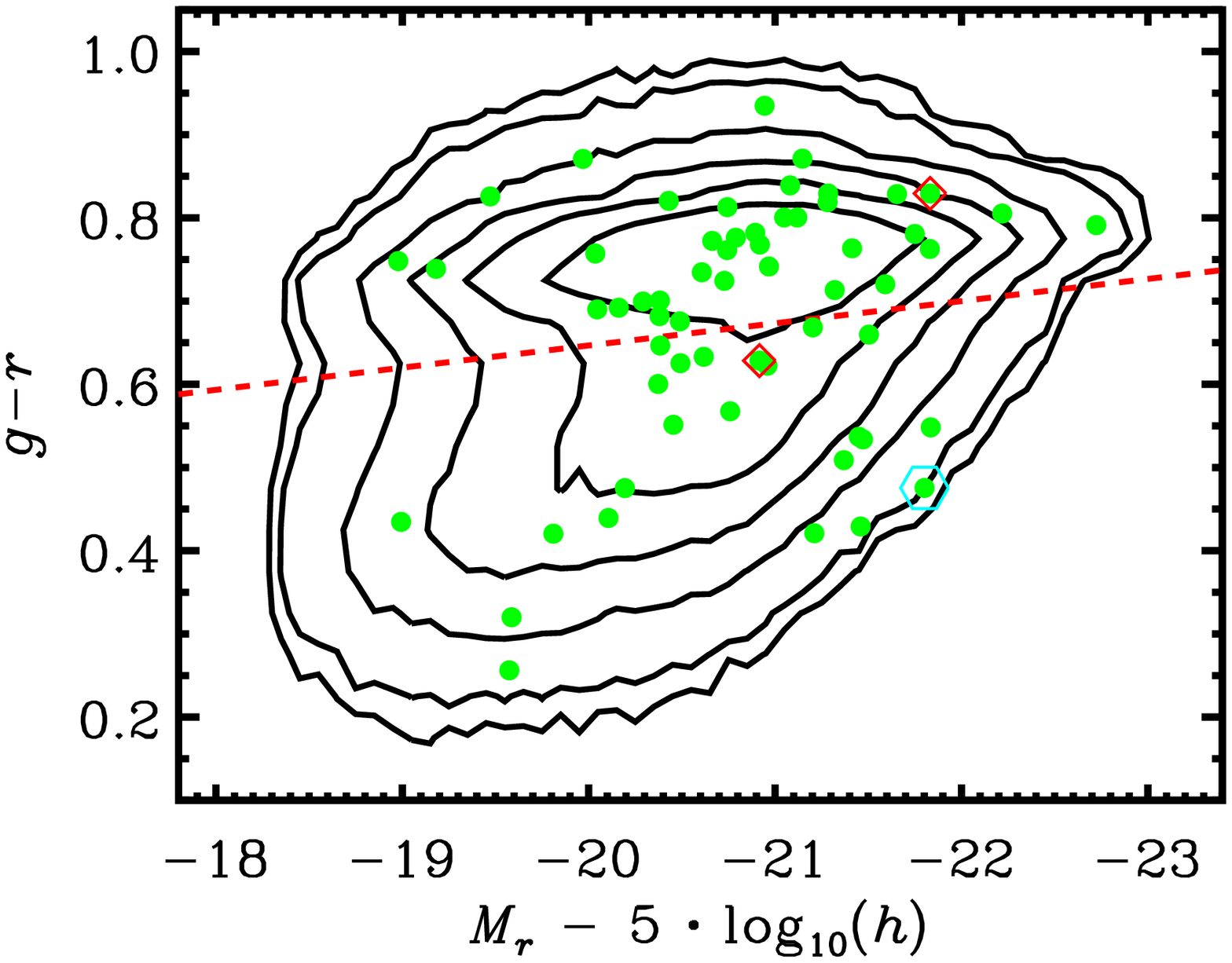}
\caption{(\emph{Left}) We plot the rest--frame $g-r$ versus $M_r$
  color--magnitude distribution for SDSS galaxies in the spectroscopic
  sample within the redshift range $0.05 < z < 0.15$. Due to the large
  number of galaxies in the sample, we plot contours (rather than
  individual points) corresponding to 25, 50, 200, 500, 1000, and 2000
  galaxies per bin of $\Delta (g-r) = 0.05$ and $\Delta M_r =
  0.1$. The dashed red horizontal line shows the division between the
  red sequence and the blue cloud as given in Equation \ref{eqn1}. The
  green points correspond to the $48$ SN Ia host galaxies included in
  sample SN--B, where the two galaxies within $1\: h^{-1}$ comoving
  Mpc of a survey edge are outlined by red diamonds. The cyan hexagon
  identifies one particular host, which is a low--redshift QSO. The
  blue circles outlining three of the green points denote the three
  SNe with ambiguous host identifications. The magenta stars
  (connected by dotted lines to the corresponding blue circle)
  illustrate the location in color--magnitude space of the other
  potential hosts. In one case, there are three possible hosts in the
  spectroscopic sample. Finally, note that the SN hosts are divided
  roughly evenly between the red sequence and the blue
  cloud. (\emph{Right}) Same as plotted on the left, except that the
  green points show the rest--frame $g-r$ versus $M_r$
  color--magnitude distribution for the $60$ SN Ia host galaxies in
  sample SN--C. Again, the two galaxies within $1\: h^{-1}$ comoving
  Mpc of a survey edge are outlined by red diamonds and the one SN
  hosted by a QSO is marked by the cyan hexagon.}
\label{cmd_fig}
\end{figure*}

As shown by many previous studies at low and intermediate redshift
\citep[e.g.,][]{strateva01, bell04, willmer06} and as illustrated in
Figure \ref{cmd_fig}, the distribution of galaxies in color--magnitude
space is bimodal, with a tight red sequence and a more diffuse blue
cloud. To divide the SDSS galaxy sample into these two broad classes,
we use the following magnitude--dependent cut:
\begin{equation}
g - r = -0.02667 \cdot M_r + 0.11333 .
\label{eqn1}
\end{equation}
This division in rest--frame $g-r$ color is shown in Fig.\
\ref{cmd_fig} as the dashed red line.

Stellar masses are computed for each galaxy in the SDSS sample, again
using the KCORRECT package with template spectral energy distributions
(SEDs) based on those of \citet{bc03}. To estimate stellar masses, the
best--fitting SED given the observed $ugriz$ photometry and
spectroscopic redshift is used to directly compute the stellar
mass--to--light ratio $({\rm M}_{*}/L)$, assuming a
\citet{chabrier03} initial mass function.

Due to the large wavelength coverage of the SDSS spectra
($3800$--$9200$\AA), star--formation rates (SFRs) are able to be
estimated for nearly every galaxy in the spectroscopic catalog based
on the measured H$\alpha$ $(\lambda6563{\rm \AA})$ line emission. For each
galaxy in the spectroscopic sample, emission--line fluxes and
equivalent widths (EWs) are measured by fitting for and subtracting
the stellar continuum, as detailed by \citet{yan06}. To derive the
line luminosity and correct for aperture effects related to the finite
size of the SDSS fibers, we estimate the total H$\alpha$ luminosity by
combining measurements of the H$\alpha$ EW with the $K$--corrected
broad--band absolute magnitudes (i.e., assuming that the ratio of the
line emission to the broad--band flux is uniform across the entire
galaxy).

Star--formation rates are inferred from the measured H$\alpha$
luminosities according to the relation given by \citet{kennicutt98}: 
\begin{equation}
\psi({\rm H}\alpha) = 7.9 \times 10^{-42} \frac{L({\rm H}\alpha)}{{\rm
  ergs}\ {{\rm s}^{-1}}} {\rm M}_{\sun}\ {\rm yr}^{-1},
\label{eqn_kenn98}
\end{equation}
where $L({\rm H}\alpha)$ is corrected by a factor of 2.8 to account
for underlying dust attenuation and stellar absorption. In a small
percentage of cases $(< \! 15\%)$, the H$\alpha$ flux is unable to be
measured accurately due to bad pixels within the emission--line or
continuum windows. In these instances, we infer the SFR from the
measured [O II] $\lambda 3727$\AA\ line luminosity, corrected using
the empirical calibration of \citet{moustakas06}. Note that in
computing our star--formation rates, we adopt a Hubble parameter, $h =
0.7$, to match that calibration.

Our measured SFRs agree well with those measured for SDSS galaxies by
\citet{brinchmann04}, who used fits of emission--line and continuum
properties to stellar population models \citep[see also][]{charlot02};
through direct comparison with our inferred star--formation rates, we
find an offset of $\sim \! 0.3$ dex and a scatter of $\sim \! 0.15$
dex relative to the \citet{brinchmann04} measurements, with the offset
of the \citet{brinchmann04} SFRs to higher values largely due to
differences in dust corrections. Direct comparison of our estimated
H$\alpha$ luminosities to those of \citet{moustakas06} show excellent
agreement, with only a small offset corresponding to a $\sim \! 0.01$
dex offset and $\sim \! 0.02$ dex scatter in the inferred
star--formation rates.

In addition to the SDSS spectroscopic data set, we utilize the larger
SDSS DR6 photometric catalog, which contains uniform, precision
photometry for millions of sources down to a $5$--$\sigma$ limiting
magnitude of $r = 22.2$ in asinh magnitudes \citep{lupton99}. For all
sources in the imaging catalog, we estimate rest--frame colors,
absolute magnitudes, stellar masses, and photometric redshifts using
the KCORRECT package. Due to the lack of spectral information, we are
unable to estimate star--formation rates for sources in the imaging
catalog.

\subsection{Identifying Host Galaxies}
\label{sec_data_hosts}

For each supernova in the SDSS--II Supernova Survey sample, we attempt
to identify a host galaxy within the SDSS DR6 spectroscopic and
imaging data sets. When identifying hosts in the SDSS spectroscopic
galaxy catalog, a projected, radial window of $25\: h^{-1}$ kpc
(physical) is employed to distinguish potential hosts on the plane of
the sky in conjunction with a velocity window of $\Delta v = 3000\
{\rm km}\ {\rm s}^{-1}$ along the line of sight. This
moderately--large velocity window is adopted to account for the
relatively--low precision of the redshifts derived from SN Ia features
\citep[$\Delta z < 0.005$,][]{frieman08}. If multiple SDSS galaxies
fall within the radial window, then the galaxy closest in projected
distance is taken as the host. A host is considered to be
unambiguously identified, if one (and only one) galaxy is found within
the search window.

In the redshift range $0.05 < z < 0.15$, a total of 48 type Ia
supernovae (SN--B sample) are matched to host galaxies (45 of the 48
are matched unambiguously) in the SDSS spectroscopic catalog. The
location of the host galaxies in color--magnitude space is given by
the green points in Figure \ref{cmd_fig}a; the hosts are roughly
equally divided between the red sequence $(26/48)$ and the blue cloud
$(22/48)$. We exclude one supernova from the SN--B sample due to its
occurrence in a bright, nearby QSO (see the hexagon point in Fig.\
\ref{cmd_fig}), for which measurements of luminosity and
star--formation rate are highly uncertain.

Although the SDSS spectroscopic data set supplies relatively precise
information about the line--of--sight position of many galaxies in the
area surveyed by the SDSS--II Supernova Survey, the SDSS imaging
catalog provides a significantly more complete census of the galaxy
population due to its much greater depth; the SDSS imaging catalog is
complete down to $r = 22.2$, while the SDSS spectroscopic sample is
magnitude--limited at $r \le 17.77$. For this reason, we also search
for host galaxies within the SDSS DR6 imaging catalog.

Using a magnitude--limited $(r \le 19.5)$ imaging catalog,\footnote{We
  employ a relatively bright magnitude limit, to ensure
  high--precision photometry and photometric redshifts, thereby
  minimizing contamination by other objects along the
  line--of--sight.} we identify potential hosts on the plane of the
sky within a projected, circular window of $25\: h^{-1}$ kpc
(physical) in radius. To differentiate between potential hosts along
the line--of--sight, we compute the redshift for each galaxy in the
imaging catalog, within the radial window, using the
photometric--redshift code \emph{SDSS\_KPHOTOZ} in KCORRECT
\citep[version v4\_1\_4][]{blanton03}. Within a velocity window of
$\Delta v = 6000\ {\rm km}\ {\rm s}^{-1}$, we select the closest
galaxy in projected distance as the host. A total of 60 type Ia
supernovae in the redshift range $0.05 < z < 0.15$ (SN--C sample) are
matched to host galaxies in the SDSS imaging catalog. The distribution
of these hosts in color--magnitude space is given by the green points
in Figure \ref{cmd_fig}b; they are weighted more towards the
red--sequence population relative to the hosts of the SN--B sample (36
out of 57 are red). The SN--C sample is a superset of the SN--B
sample, with all of the supernovae in SN--B being matched to the same
host galaxy. Again, we exclude the one supernova from the sample that
is found within a local QSO.

\subsection{Measuring the Local Environment}
\label{sec_data_environ}
We consider the ``environment'' of a galaxy to be defined by the local
mass overdensity, as traced by the local overdensity of galaxies; over
quasi--linear regimes, the mass density and galaxy density should
simply differ by a factor of the galaxy bias \citep{kaiser87}. To
estimate the overdensity of galaxies in the SDSS, we utilize
measurements of the projected fifth--nearest--neighbor surface density
$(\Sigma_5)$ about each galaxy, where the surface density depends on
the projected distance to the fifth--nearest neighbor, $D_{p,5}$, as
$\Sigma_5 = 5 / (\pi D^{2}_{p,5})$. In computing $\Sigma_5$, a
velocity window of $\pm1500\ {\rm km}\ {\rm s}^{-1}$ is employed to
exclude foreground and background galaxies along the line of sight.
The projected distance to the N$^{\rm th}$--nearest neighbor provides
an accurate estimate of local galaxy density over a broad and
continuous range of scales. As shown by \citet{cooper05}, it is
reasonably robust to redshift--space distortions, while also
effectively tracing the local density in underdense regions.

To correct for the redshift dependence of the sampling rate of the
SDSS, each surface density value is divided by the median $\Sigma_5$
of galaxies at that redshift within a window of $\Delta z = 0.02$;
this converts the $\Sigma_5$ values into measures of overdensity
relative to the median density (given by the notation $1 + \delta_5$
herein) and effectively accounts for redshift variations in the
selection rate \citep{cooper05}. We restrict our analyses to the
redshift range $0.04 < z < 0.16$, avoiding the low-- and
high--redshift tails of the SDSS $d{\rm N}/dz$ distribution where the
variations in the survey selection rate are greatest. Finally, to
minimize the effects of edges and holes in the SDSS survey geometry,
we exclude all galaxies within $1\: h^{-1}$ Mpc (comoving) of a survey
boundary, reducing our sample size to $392,938$ galaxies within the
redshift range $0.04 < z < 0.16$.

For each supernova (those with and without an identified host in the
SDSS DR6 galaxy catalog), we measure the local environment in a manner
identical to that followed for the galaxy sample. That is, we measure
the local surface density of galaxies about the position of the
supernova, using the positional information $(\alpha, \delta, z)$ from
the SDSS--II Supernova Survey and using the SDSS DR6 galaxy sample to
trace the local environment. After excluding SNe near the survey
boundary (within $1\: h^{-1}$ Mpc), we arrive at a final sample
(SN--A) of $134$ SNe Ia. Note that for the subset of SNe with an
identified host in the SDSS DR6 spectroscopic catalog (SN--B),
higher--precision information about the local environment is also
available by proxy via the host galaxy. A summary of all of the
supernova samples is provided in Table \ref{data_sample_tab}.

\subsection{Selecting Comparison Samples}
\label{sec_data_samples}
As discussed in \S \ref{sec_intro}, a variety of recent observations
have shown that the SN Ia rate (weighted by mass or by luminosity)
depends on the properties of the host galaxy. For instance,
\citet{mannucci05}, using the supernova catalog of
\citet{cappellaro99}, showed that SNe Ia are more common in
morphologically late--type galaxies (Irr and Sbc/d) relative to more
bulge--dominated systems (E/S0). Similarly, recent work from the
Supernova Legacy Survey \citep{sullivan06a} found that the SN Ia rate
is greater among galaxies with greater stellar mass and among galaxies
with higher star--formation rates \citep{sullivan06b}.

For several decades, the observed properties of galaxies (including
star--formation rates, morphology, and rest--frame color) have been
known to depend upon the local environment \citep[e.g.,][]{davis76,
  postman84, balogh98, cooper06}. In particular, galaxies with more
massive stellar populations tend to favor regions of higher galaxy
density \citep{hogg04, zehavi05}, while systems with high
star--formation rates typically reside in low--density environs in the
local Universe \citep{gomez03, cooper08a}. In addition, the
relationships between galaxy properties and environment depend on
redshift, with the color--density and morphology--density relations
growing weaker at higher $z$ \citep[e.g.,][]{dressler97, smith05,
  cooper07} and the SFR--density relation inverting between $z \sim 1$
and $z \sim 0$ \citep{elbaz07, cooper08a}.

\begin{figure}[h]
\centering
\plotone{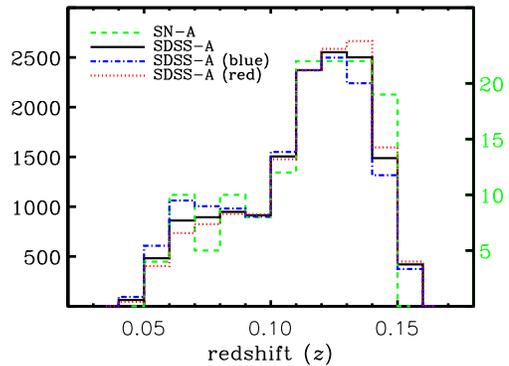}
\caption{The redshift distributions for the $134$ SNe Ia in sample
  SN--A and for the $15,000$ galaxies in sample SDSS--A. As summarized
  in Table \ref{data_sample_tab}, the SDSS--A galaxy sample is
  randomly chosen to match the redshift distribution of the
  supernovae. The red dotted and blue dashed--dotted lines show the
  redshift distributions for red and blue subsets of the SDSS--A
  sample, respectively, following the division in rest--frame $g-r$
  color as given by Equation \ref{eqn1}. Note that all histograms are
  normalized to have equal area. The total number of objects per bin
  for the SDSS--A and SN--A samples is shown by the left and right
  axis scales, respectively.}
\label{zdist1_fig}
\end{figure}

Given these known correlations between galaxy properties and [1] the
SN Ia rate as well as [2] the local galaxy density, we extract
multiple subsamples from the SDSS galaxy sample, selected to match the
characteristics of the SN Ia samples, thereby enabling analysis of the
local environments of SN hosts independent of selection biases
connected to galaxy type. Since many of the SNe in the SN--A sample
lack an identified host galaxy in the SDSS galaxy catalog, we are
unable to select a comparison sample that matches properties such as
luminosity, stellar mass, star--formation rate, etc. Here, we randomly
select $15,000$ galaxies from the set of $392,938$ SDSS galaxies with
accurate environment measures, so as to match their redshift
distribution to that of the SN--A sample. We utilize a matching radius
of $\Delta z = 0.01$, randomly drawing galaxies from the redshift
range $0.04 < z < 0.16$. In Fig.\ \ref{zdist1_fig}, we show the
redshift distributions for this subsample (SDSS--A) relative to that
of the SNe in the SN--A sample.

For the smaller SN--B sample, however, we are able to randomly select
a comparison sample that matches galaxy properties such as luminosity,
color, and stellar mass. From the set of $392,938$ SDSS galaxies with
accurate environment measures, we draw two comparison samples: one
matched to the luminosity, color, and redshift distributions of the
SN hosts in sample SN--B (sample SDSS--B) and a second matched to the
stellar mass, star--formation rate, and redshift distributions of the
same host galaxies (sample SDSS--F). 

Members of the comparison samples are drawn randomly from within
3--dimensional radial windows of $\Delta(g-r)^2 + \Delta M_r^2 +
\Delta z^2 < 0.002$ and $\Delta \log(\psi)^2 + \Delta \log({\rm
  M}_{*})^2 + \Delta z^2 < 0.002$, centered on the properties of each
host. The SDSS--B and SDSS--F samples are constructed from $7,500$
independent, random matches. Thus, some SDSS galaxies are duplicated
in the comparison samples; however, the large size of the SDSS
spectroscopic galaxy catalog ensures that duplication is minimized
such that $> \! 80\%$ of the comparison samples are unique, with no
individual galaxy included more than $8$ times in a given comparison
sample.

\begin{figure}[h]
\centering
\plotone{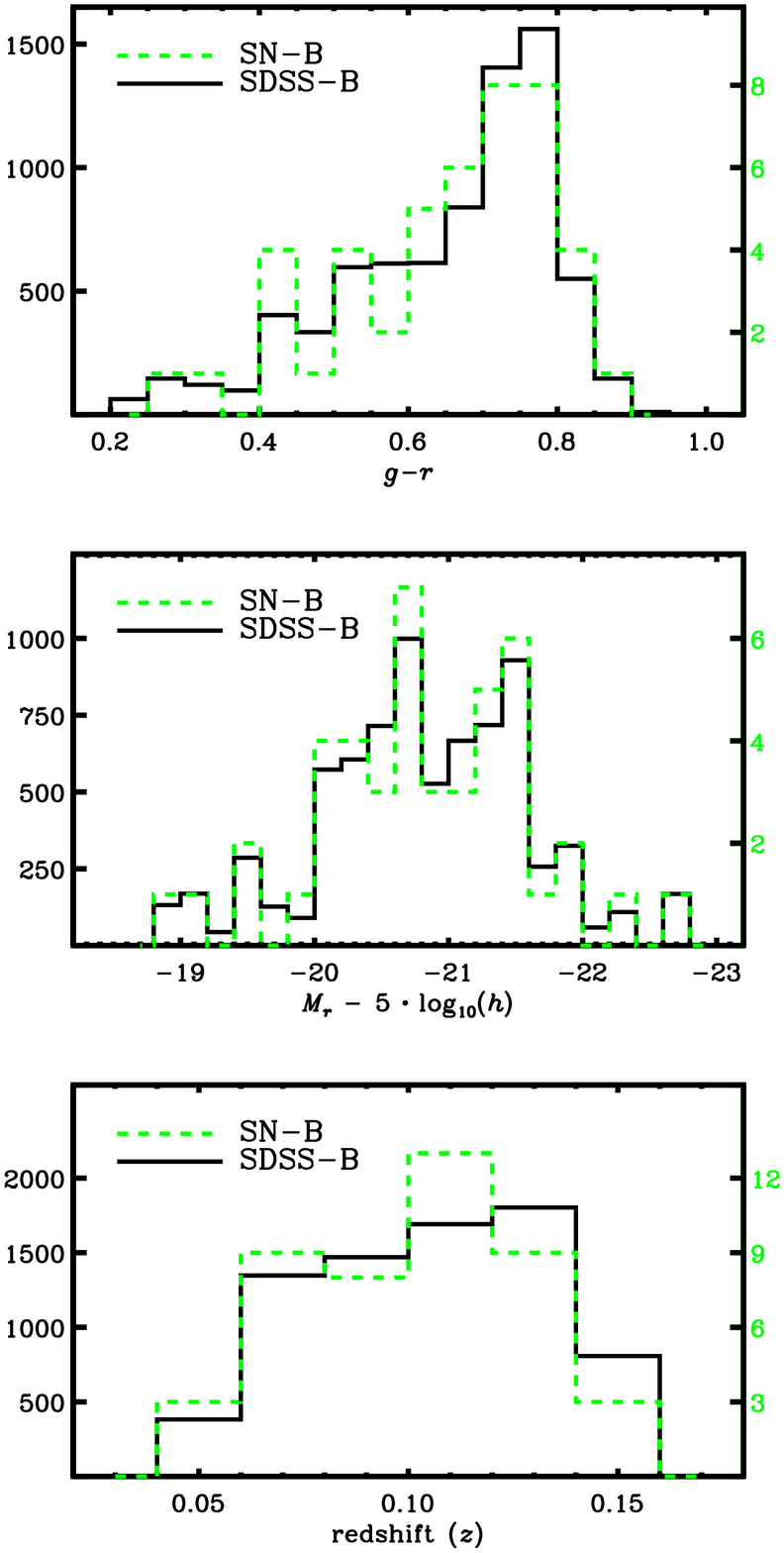}
\caption{The rest--frame color $(g-r)$, absolute magnitude $(M_r)$,
  and redshift $(z)$ distributions for the $45$ SNe Ia host galaxies
  in sample SN--B and for the $7,500$ galaxies in sample SDSS--B. All
  histograms are normalized to have equal area. The total number of
  objects per bin for the SDSS--B and SN--B samples is shown by the
  left and right axis scales, respectively. As summarized in Table
  \ref{data_sample_tab}, the SDSS--B galaxy sample is randomly chosen
  to match the color, luminosity, and redshift distributions of the
  supernovae hosts.}
\label{cmzdist_fig}
\end{figure}

\begin{figure*}[tb]
\centering
\plottwo{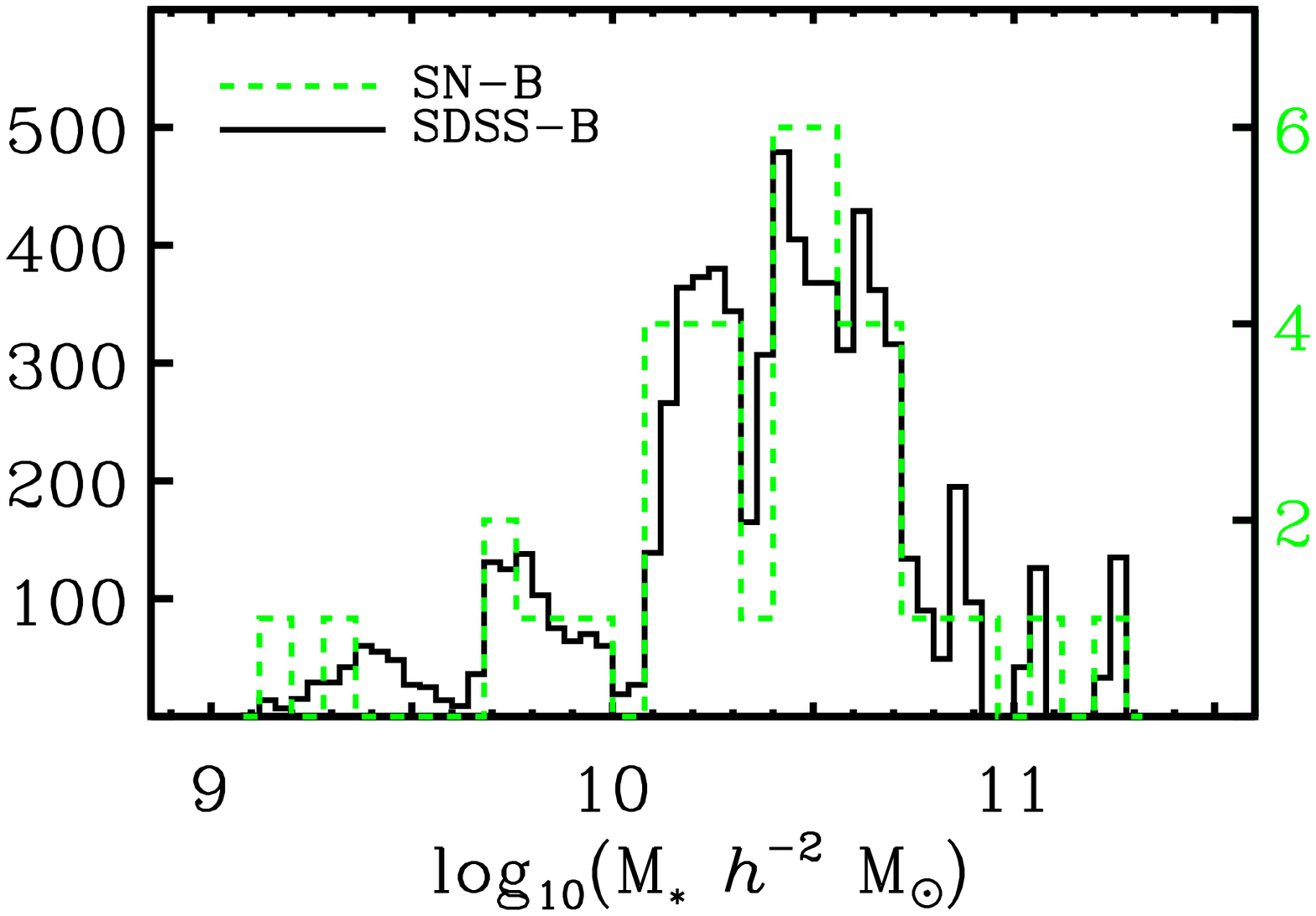}{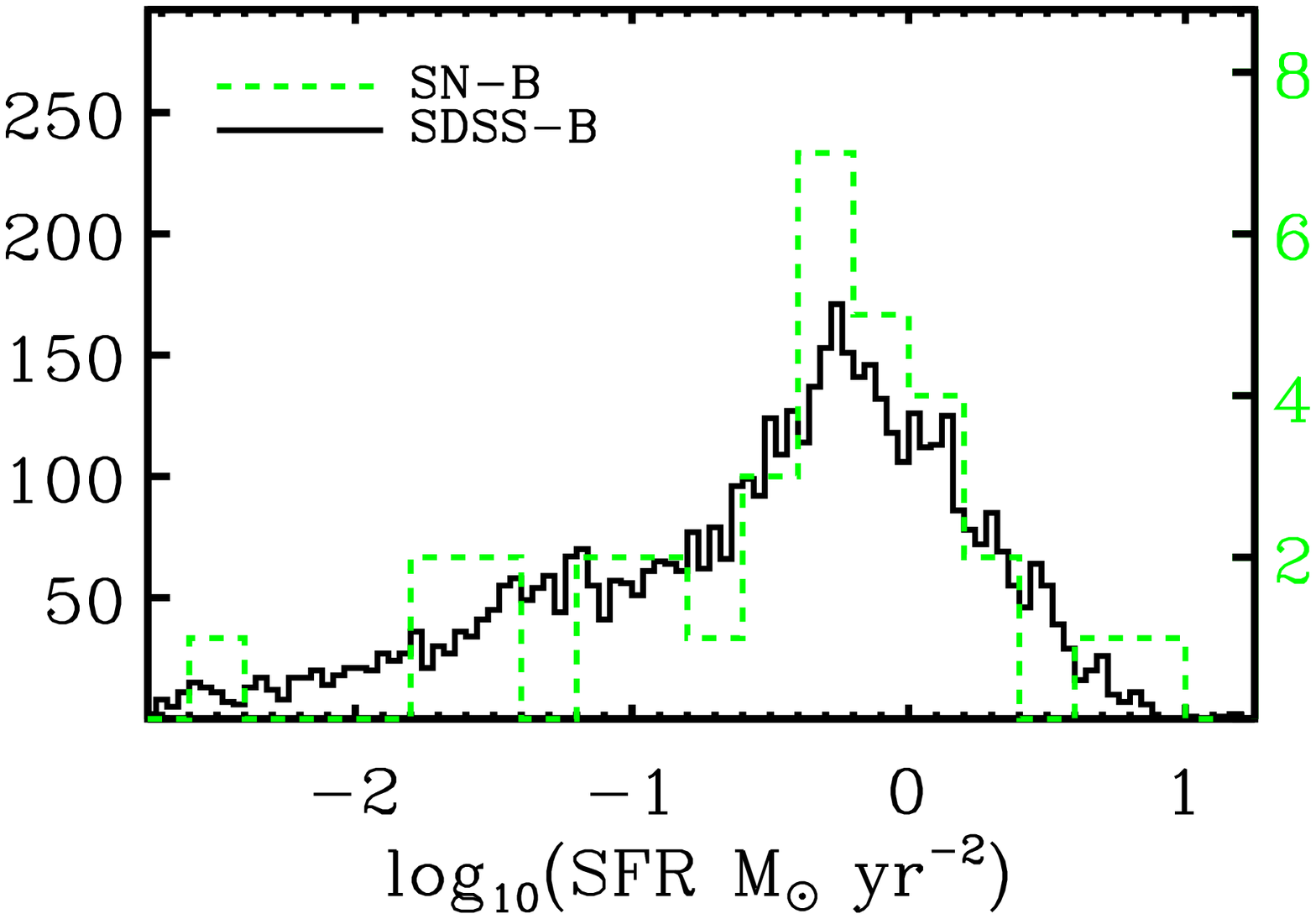}
\caption{The distribution of stellar masses (\emph{left}) and
  star--formation rates (\emph{right}) for the $47$ SNe Ia host
  galaxies in sample SN--B and for the $7,500$ galaxies in sample
  SDSS--B. Each histogram is normalized to have equal area, with the
  total number of objects per bin for the SDSS--B and SN--B samples
  indicated by the left and right axis scales, respectively. As
  summarized in Table \ref{data_sample_tab}, the SDSS--B galaxy sample
  is randomly chosen to match the color, luminosity, and redshift
  distributions of the supernovae hosts; however, the stellar mass and
  star--formation rate distributions are well matched too. This
  reflects the strong correlations between specific SFR and stellar
  mass--to--light ratio with rest--frame color.}
\label{mass_fig}
\end{figure*}

Figure \ref{cmzdist_fig} shows the relative distributions of
rest--frame color, luminosity, and redshift for the galaxies in the
SN--B and SDSS--B samples; by design, the distributions of these
properties are well matched. Given the relatively tight relationship
between the combination of rest--frame optical color and luminosity
with stellar mass and star--formation rate
\citep[e.g.,][]{kauffmann03, cooper08a}, we find that the
distributions of stellar masses and star--formation rates for the
SN--B and SDSS--B samples are also closely matched (see Figure
\ref{mass_fig}). The SDSS--F sample, which is directly matched to the
stellar mass and star--formation rates of the host galaxies in the
SN--B sample, shows similar distributions of these galaxy properties.

Finally, we define a comparison sample (SDSS--C), selected to match
the distribution of rest--frame colors, luminosities, and redshifts of
the SN--C supernova sample. Recall that the SN--C sample is selected
by matching the SDSS--II supernovae to the SDSS DR6 imaging
catalog. Due to the lack of spectroscopic information for all of the
hosts, we are unable to estimate accurate star--formation rates for
the SN--C sample. The details of each of the supernova and galaxy
samples is given in Table \ref{data_sample_tab}.

\begin{deluxetable*}{l l l l l}
\tablewidth{0pt}
\tablecolumns{4}
\tablecaption{\label{data_sample_tab} Descriptions of the SNe Ia and
  Comparison Galaxy Samples} 
\tablehead{ Sample & $N_{\rm obj}$ & $N_{\rm edge-cut}$ & $z$ range & Brief Description }
\startdata
SN--A & 163 & 134 & $0.05 < z < 0.15$ & 
\parbox[l]{2.75in}{all SNe Ia with spectra from the SDSS--II SN Survey} \\
&  &  &  & \\
SDSS--A & 15,000 & 15,000 & $0.04 < z < 0.16$ &
\parbox[l]{2.75in}{galaxies chosen randomly to match the redshift
  distribution of the SN--A sample} \\
&  &  &  & \\
SN--B & 48 & 45 & $0.05 < z < 0.15$ &
\parbox[l]{2.75in}{all SNe Ia with spectra from the SDSS--II SN Survey
  and with identified host galaxies within $R < 25\ h^{-1}$ kpc in the
  SDSS spectroscopic sample, selecting the closest host in projected
  distance in cases of confusion} \\
&  &  &  & \\
SDSS--B & 7,500 & 7,500 & $0.04 < z < 0.16$ &
\parbox[l]{2.75in}{galaxies chosen randomly to match the luminosity,
  color, and redshift distributions of the SN--B sample} \\
&  &  &  & \\
SN--C & 60 & 57 & $0.05 < z < 0.15$ &
\parbox[l]{2.75in}{all SNe Ia with spectra from the SDSS--II SN Survey
  and with identified host galaxies within $R < 25\ h^{-1}$ kpc in the
  SDSS imaging sample} \\ 
&  &  &  & \\
SDSS--C & 7,500 & 7,500 & $0.04 < z < 0.16$ &
\parbox[l]{2.75in}{galaxies chosen randomly to match the luminosity,
  color, and redshift distributions of the SN--C sample} \\
&  &  &  & \\
SN--D & 48 & 45 & $0.05 < z < 0.15$ &
\parbox[l]{2.75in}{all SNe Ia with spectra from the SDSS--II SN Survey
  and with identified host galaxies within $R < 25\ h^{-1}$ kpc in the
  SDSS spectroscopic sample, selecting the bluest host in cases of
  confusion} \\  
&  &  &  & \\
SDSS--D & 7,500 & 7,500 & $0.04 < z < 0.16$ &
\parbox[l]{2.75in}{galaxies chosen randomly to match the luminosity,
  color, and redshift distributions of the SN--D sample} \\
&  &  &  & \\
SN--E & 48 & 45 & $0.05 < z < 0.15$ &
\parbox[l]{2.75in}{all SNe Ia with spectra from the SDSS--II SN Survey
  and with identified host galaxies within $R < 40\ h^{-1}$ kpc in the
  SDSS spectroscopic sample, selecting the closest host in projected
  distance in cases of confusion} \\ 
&  &  &  & \\
SDSS--E & 7,500 & 7,500 & $0.04 < z < 0.16$ &
\parbox[l]{2.75in}{galaxies chosen randomly to match the luminosity,
  color, and redshift distributions of the SN--E sample} \\
&  &  &  & \\
SDSS--F & 7,500 & 7,500 & $0.04 < z < 0.16$ &
\parbox[l]{2.75in}{galaxies chosen randomly to match the stellar mass,
  star-formation rate, and redshift distributions of the SN--B sample} \\
\enddata
\tablecomments{We list each supernova and galaxy sample employed in
  the analysis, detailing the selection cut used to define the sample
  as well as the redshift range covered and the number of objects
  included before $(N_{\rm obj})$ and after $(N_{\rm edge-cut})$
  removing those within $1\: h^{-1}$ comoving Mpc of a survey edge.}
\end{deluxetable*}

\section{Results}
\label{sec_results}

\begin{figure}[h]
\centering
\plotone{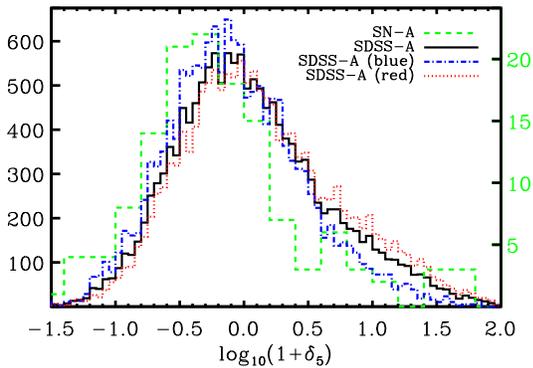}
\caption{We plot the distribution of overdensities for the SN--A
  (green dashed line) and SDSS--A (black solid line) samples. In
  addition, we divide the SDSS--A sample according to rest--frame
  $g-r$ color, following Equation \ref{eqn1}, and plot the environment
  distributions for the resulting red (red dotted line) and blue (blue
  dashed-dotted line) subsamples. All of the histograms are normalized
  to have equal area, with the total number of objects per bin for the
  SDSS--A and SN--A samples indicated by the left and right axis
  scales, respectively. We find that the environment distribution for
  the SNe Ia in sample SN--A is skewed to lower overdensities relative
  to a random sample of galaxies from the SDSS with a similar redshift
  distribution.}
\label{envdistA_fig}
\end{figure}

In Figure \ref{envdistA_fig}, we show the environment distribution for
the 134 supernovae in the SN--A sample in comparison to that of the
SDSS--A galaxy sample. The SNe appear to be clearly biased towards
lower--density environs relative to a sample of random galaxies with
the same redshift distribution. Moreover, the overdensity distribution
for the supernovae looks to be skewed towards low overdensities even
relative to that of the blue galaxies in SDSS--A, where the blue
subsample is selected following the rest--frame color division given
by Equation \ref{eqn1}.

A variety of statistical tests have been developed to determine
whether two sets of data come from the same underlying distribution.
We have applied two of the most powerful non--parametric tests (i.e.,
those which are independent of Gaussian assumptions) to our data: the
one--sided Wilcoxon--Mann--Whitney U test \citep{mann47, press92,
  wall03} and the Kolmogorov--Smirnov test \citep{press92,
  wall03}. The result of each test is a $P$--value: the probability
that a value of the U or K--S statistic equal to the observed value or
more extreme would be obtained, if some ``null'' hypothesis
holds. Throughout the remainder of this paper, results with a
$P$--value below 0.05 (corresponding closely to 2$\sigma$ for a
Gaussian) will be considered to be ``significant'', while $P$--values
below 0.01 will be classified as ``highly significant''.

The Wilcoxon--Mann--Whitney U statistic is computed by ranking all
elements of the two datasets together, and then comparing the mean (or
total) of the ranks from each dataset. Because it relies on ranks,
rather than observed values, it is highly robust to non--Gaussianity,
but still has efficiency (as measured by the sample size required to
reach a given error level) almost as high as the classical $t$ test
for Gaussian distributions. In particular, we apply a one--sided U
test, which determines the $P$--value for the null hypothesis that a
specified sample is not skewed to higher values than the other sample;
the obvious differences between the environments of the blue supernova
hosts and blue galaxy samples (e.g., see Figure \ref{envdistB2_fig})
make a two--sided test (for the null hypothesis that the samples have
the same distribution in either sense) less appropriate. Since the
test is one--sided, possible $P$--values range from $0$ to $0.5$. The
Wilcoxon--Mann--Whitney (WMW) U test is particularly useful for small
datasets (as we have for our SN samples) due to its insensitivity to
outlying data points, its avoidance of binning, and its high
efficiency.

In addition to the U test, we also employ the two--sided
Kolmogorov--Smirnov (KS) test. This test quantifies differences
between samples using the maximum absolute difference between the
cumulative distributions of two datasets, the K--S statistic. This
difference will be small for data drawn from the same distribution and
large for dissimilar data; it is particularly sensitive to differences
in the ``core'' of a distribution, but relatively insensitive to
differences in tails, lending the test high robustness to
non--Gaussianity. The $P$--value for the test is then the probability
of obtaining the observed value of the K--S statistic, or a higher one,
for the null hypothesis that we have two random datasets drawn from
the same underlying distribution. If this probability is small, then
the likelihood is high that the distributions of the data are not
identical.

Performing a one--sided WMW U test on the overdensity measures for the
SN--A and SDSS--A samples, we find that the SN--A environment
distribution is skewed to smaller values of overdensity than that of
the SDSS--A galaxy sample, with a $P$--value $< 0.01$, i.e., there is
less than a $1\%$ chance that we would observe a difference this
strong if both samples were drawn from the same parent
distribution. If we compare to the blue SDSS--A galaxy subsample, we
still obtain $P = 0.02$, which is highly significant. If instead we
simply compute means and errors on the mean for the various
overdensity distributions in Fig.\ \ref{envdistA_fig}, we arrive at
this same general result (see Table \ref{envdist_tab}). The mean
overdensity, $<\log_{10}(1+\delta_5)>$, for the SNe in sample SN--A is
notably lower than the mean environment of the SDSS--A sample, with
the difference significant at a $> \! 3 \, \sigma$ level. Relative to
the mean overdensity of just the blue galaxies in the SDSS--A sample,
the typical environment of the supernovae is still underdense, but
with the difference only significant at a $\sim \! 1 \, \sigma$
level. A KS test confirms these results, indicating that the SN--A and
SDSS--A environment distributions are very unlikely to have been drawn
from the same parent distribution.

The differences in the environment distributions for the SN--A and
SDSS--A samples, however, are potentially only driven by differences
in the properties of the two galaxy samples in concert with the
underlying correlations between galaxy properties (e.g., color,
morphology, etc.) and environment. For instance, a sample of galaxies
chosen randomly from the SDSS DR6 galaxy catalog will undoubtedly turn
out to be dominated by red galaxies; the red fraction is $60\%$ for
the full SDSS sample at $0.05 < z < 0.15$.\footnote{The red fraction
  increases with redshift in the SDSS sample and is actually $> \!
  60\%$ at $z \sim 0.12$, the median redshift of the SDSS--A sample
  (see Fig.\ \ref{zdist1_fig}).}  If we assume that the hosts of the
SNe in the SN--A sample have colors that relatively closely follow the
distribution of the SN--B hosts (roughly evenly divided between red
and blue), then our SDSS--A sample is likely somewhat more skewed
towards red galaxies than are the hosts of the SN--A sample. Since
galaxy color is strongly correlated with environment, with red
galaxies favoring overdense regions \citep[e.g.,][]{blanton05a}, the
SDSS--A sample is potentially biased towards a sample of galaxies in
higher--density regions, compared to the SN hosts, just because of
this effect. For this reason, meaningful conclusions cannot be drawn
from the analysis of the SN--A and SDSS--A samples.

The SN--B sample, which includes only those SNe with identified host
galaxies in the SDSS spectroscopic sample, allows us to compare the
environments of type Ia supernovae to the environments of a sample of
galaxies with similar properties to those of the SN hosts, thereby
avoiding the biases that weaken the utility of the larger SN--A
sample. As discussed in \S \ref{sec_data_samples}, the SN--B sample
has the additional benefit of more precise environment estimates for
the supernovae (relative to the estimates for the SN--A sample), since
the host galaxy redshifts have roughly $50$ times higher precision
than the supernova redshifts.

\begin{figure}[h]
\centering
\plotone{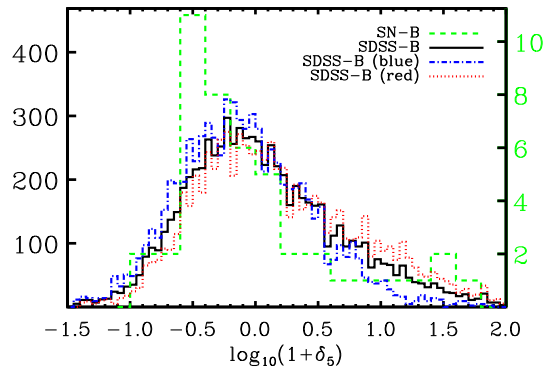}
\caption{We plot the distribution of overdensities for the SN--B
  (green dashed line) and SDSS--B (black solid line) samples. In
  addition, we divide the SDSS--B sample according to rest--frame
  $g-r$ color, following Equation \ref{eqn1}, and plot the environment
  distributions for those red (red dotted line) and blue (blue
  dashed-dotted line) subsamples. All of the histograms are normalized
  to have equal area, with the total number of objects per bin for the
  SDSS--B and SN--B samples indicated by the left and right axis
  scales, respectively. We find that the environment distribution for
  the type Ia SN hosts in the SN--B sample is skewed to lower
  overdensities relative to a randomly--selected sample of galaxies
  with matching luminosity, color, and redshift distributions.}
\label{envdistB_fig}
\end{figure}

\begin{figure*}[tb]
\centering
\plottwo{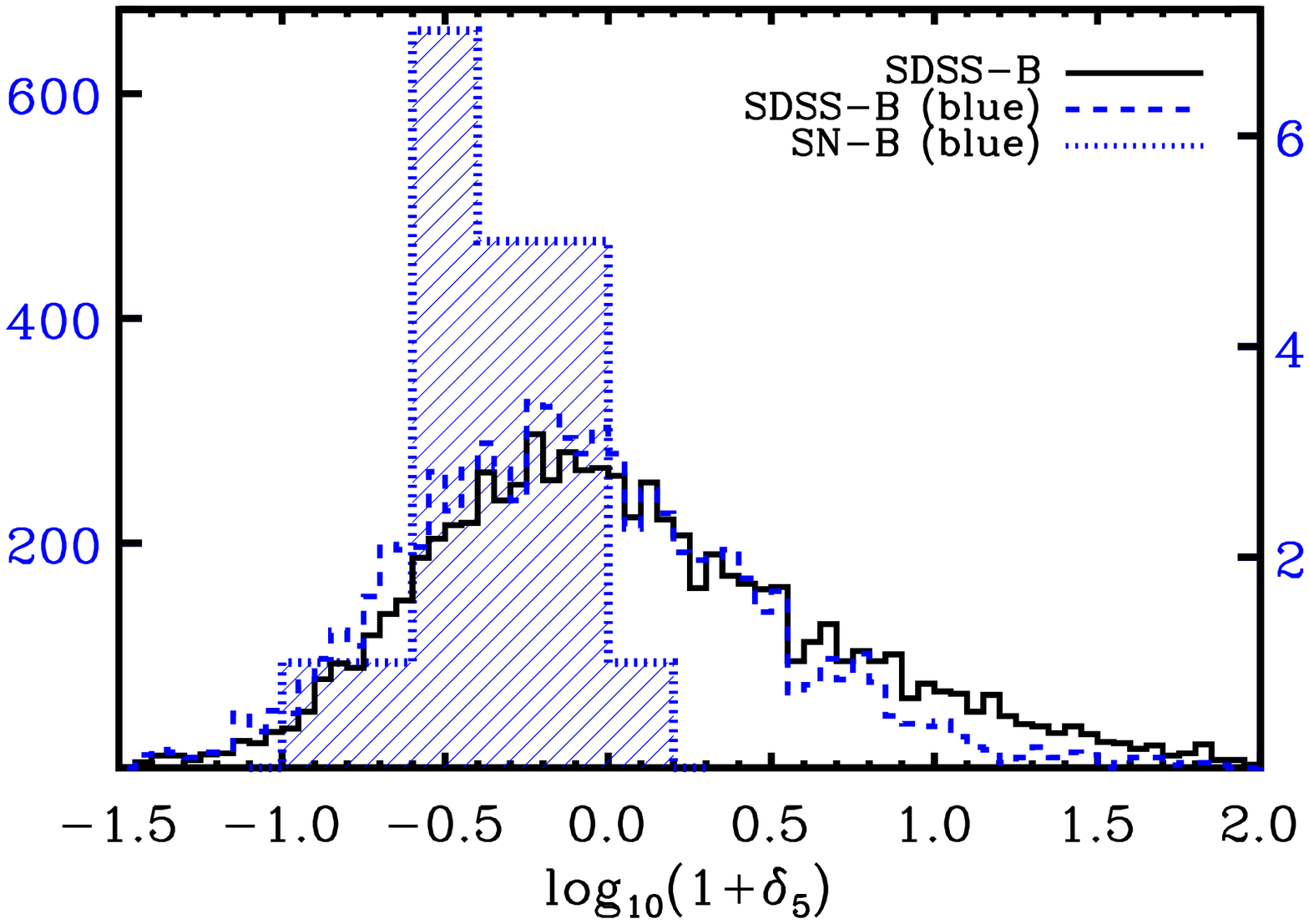}{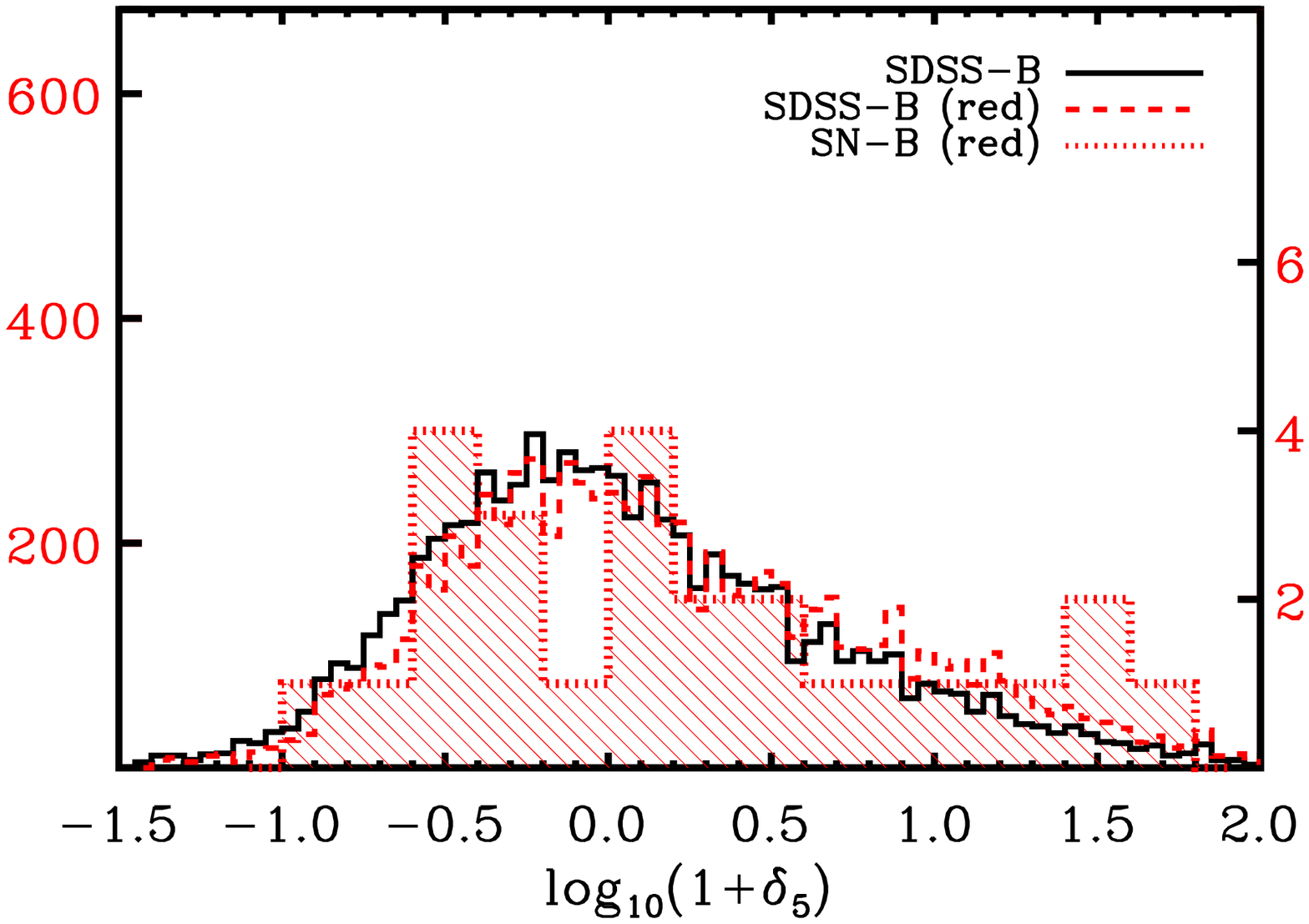}
\caption{We plot the distribution of overdensities for the blue
  (\emph{left}) and red (\emph{right}) galaxies in the SN--B and
  SDSS--B samples. All of the histograms are normalized to have equal
  area. The total number of objects per bin for the blue or red
  SDSS--B and SN--B samples is shown by the left and right axis
  scales, respectively. We find that the environment distribution for
  the blue type Ia SN hosts in the SN--B sample is skewed to lower
  overdensities relative to a sample of blue galaxies with similar
  luminosity, color, and redshift distributions.}
\label{envdistB2_fig}
\end{figure*}

In Figure \ref{envdistB_fig}, we show the environment distribution for
the 45 SNe Ia in the SN--B sample alongside that for the SDSS--B
sample, which was selected to match the color, luminosity, and
redshift distributions of the 45 SN hosts. As we found for the SN--A
sample, the SNe Ia in the SN--B sample appear to be more commonly
found in lower--density regions relative to both the blue and the red
galaxies in the SDSS--B sample. Performing a WMW U test on the
overdensity measures for the SN--B and SDSS--B samples confirms that
the SN--B environment distribution is distinct from that of the
SDSS--B galaxy sample with a $P$--value less than $0.04$, a
significant result.

By dividing the SN--B sample according to the rest--frame color of the
host galaxy (again, following Equation \ref{eqn1}), we are also able
to directly compare the environment distributions for the blue and red
galaxies separately. In Figure \ref{envdistB2_fig}, we plot the
distributions of overdensity measures for these subsamples selected by
rest--frame color. While the distributions for the red galaxies in
SN--B and SDSS--B appear to be quite similar, the overdensities for
the blue SN Ia hosts are significantly skewed to lower values relative
to the blue galaxies in the SDSS--B sample. Wilcoxon--Mann--Whitney U
tests confirm these impressions, concluding that the overdensity
measures for the blue hosts are stochastically smaller than those of
the blue SDSS--B galaxies at a $> \! 99\%$ level, while the
distributions for the red SN--B and SDSS--B galaxy samples are
indistinguishable from each other. Examining the mean overdensities,
$<\log_{10}(1+\delta_5)>$, for the respective samples confirms the
result of the WMW U tests. As shown in Table \ref{envdist_tab}, the
mean environments for the blue samples differ at a $> \! 4 \, \sigma$
level, with the SN hosts being typically found in lower--density
regions relative to galaxies of like color, luminosity, and
redshift. Two--sided Kolmogorov--Smirnov (KS) tests similarly find
that the blue hosts in the SN--B sample have an environment
distribution distinct from that of the blue SDSS--B galaxies with $P <
0.01$, while no significant distinction is found when comparing the
environments of the red hosts in SN--B and the red galaxies in
SDSS--B.

The difference between the environments of the host galaxies in the
SN--B sample and the environments of like galaxies is further
supported by comparing the overdensity measures for SN--B to those of
the SDSS--F sample, which is a galaxy sample selected to match the
stellar masses, SFRs, and redshifts of the SN--B host galaxies, rather
than their colors, luminosities, and redshifts. As shown in Table
\ref{envdist_tab} and Table \ref{envdist2_tab}, we find that the blue
SN hosts in SN--B are biased to lower--density regions relative to the
blue galaxies in the SDSS--F sample, with a $P$--value (or
significance) similar to that found when comparing SN--B to SDSS--B.

While the environment measures for the SN--C sample are less precise
than those of SN--B, due to the lack of a spectroscopic redshift for
each host galaxy, we still detect a significant difference between the
environment distribution of blue hosts in the SN--C sample in
comparison to like galaxies in the SDSS--C sample. The mean
overdensities for the blue subsamples differ at a $3.5 \sigma$ level,
while the typical environments of the red subsamples are
indistinguishable within the uncertainties. The WMW U and KS tests
support the results derived from analyzing the mean overdensities,
with the environment distribution for the blue hosts distinguishable
from that of a color--, luminosity--, and redshift--matched sample at
$P < 0.01$ and $P < 0.03$, respectively.

Altogether, the primary result of analyzing the large--scale
environments of samples SN--B and SN--C is that blue SN Ia host
galaxies are found to be biased towards low--density environments
relative to galaxies of like stellar mass, star--formation rate, and
redshift. As discussed in more detail in \S \ref{sec_disc}, this
result can be interpreted as evidence for a bias in the rate \emph{or}
luminosity of type Ia events in low--density regions, such that prompt
Ia events are more numerous or more luminous in underdense environs.

One of the most striking features of the environment distribution for
the blue hosts in SN--B, as shown in Fig.\ \ref{envdistB_fig}, is the
complete lack of blue host galaxies in high--density regions. All of
the supernovae in star--forming systems are found in overdensities of
$\log_{10}(1+\delta_5) < 0.14$, while the blue comparison galaxies in
the SDSS--B sample span the full range of overdensities. To test the
significance of this sharp cut--off in overdensity, we draw $100,000$
random subsets of $20$ galaxies each from the blue SDSS--B sample. Of
these $100,000$ subsamples, we find that less than $0.1\%$ display
sharp cut--offs in their environment distribution such that all $20$
galaxies reside in overdensities of $\log_{10}(1+\delta_5) < 0.15$.
This analysis further supports the conclusion that the SN Ia rate or
luminosity is elevated in low--density environments (relative to more
overdense environs) among star--forming galaxies.

We test the strength (or depth) of this cut--off in overdensity by
measuring how much of an increase in the type Ia rate in low--density
environments is needed to make the observed cut--off at
$\log_{10}(1+\delta_5) = 0.15$ statistically likely to occur. To do
this, we select the $3244$ random blue galaxies in the SDSS--B sample,
which by design match the rest--frame color, absolute magnitude, and
redshift distributions of the blue SN Ia hosts in sample SN--B. From
this parent population, we then draw $100,000$ independent samples of
$20$ random galaxies with the likelihood of drawing an object in a
low--density $(\log_{10}(1+\delta_5) < 0.15)$ environment forced to be
$2\times$ greater than that of a galaxy in a high--density
$(\log_{10}(1+\delta_5) \ge 0.15)$ environment. The division between
the underdense and overdense regimes is selected to be
$\log_{10}(1+\delta_5) = 0.15$, so as to maximize the probability of
seeing an apparent cut--off. We repeat this exercise while varying the
degree to which the SN rate increases in the underdense regime (e.g.,
$3\times$ and $4\times$ greater, etc.). 

We find that to have a $1\%$, $5\%$, or $32\%$ probability of
observing a cut--off at $\log_{10}(1+\delta_5) = 0.15$ would require
an increase in the SN Ia rate in low--density environments at the
level of $> \! 1.5\times$, $2.5\times$, and $7\times$,
respectively. Thus, a SN Ia rate $\gtrsim \! 2\times$ higher within
star--forming galaxies in low--density environments (relative to those
in more overdense regions) is likely required to produce the observed
cut--off in the environment distribution among the blue SN host galaxy
population.

From this statistical analysis, we have determined the probability of
observing so strong a strong cut--off in environment, given a SN rate
in high--density environs which is $R$ times that in low--density
regions, as a function of $R$. From this, we can determine the
Bayesian equivalent of a 95\% confidence interval, a 95\% credible
interval, as this will be the region containing 95\% of the posterior
probability. Bayes' theorem \citep{press92, wall03} shows that this
posterior probability will be proportional to the probability of
obtaining a cut--off as strong as observed (or stronger) for a given
$R$ --- the likelihood --- multiplied by the probability distribution
we would assign to $R$ in the absence of any measurements --- a prior.
Based on this analysis, we conclude that there is $95\%$ probability
that $R < 0.38$; i.e., that the SN Ia rate is less than $0.38$ times
as large in the high--density regime ($\log_{10}(1+\delta_5) \ge
0.15$) as at lower densities, assuming a flat prior probability
distribution for $R$, a conventional zero--information prior for a
parameter with some characteristic scale (e.g., $R=1$). If we instead
adopt a prior with uniform probability for all intervals of $\log(R)$
(i.e., $P(R) =1/R$), we would find that there would be only a $5\%$
probability that $R>0.12$. Finally, even with a prior as extreme as
$P(R) = R$, which favors $R>1$, we find that there is $95\%$
probability that $R<0.66$; an equal SN rate in high-- and low--density
regions ($R=1$) is strongly ruled out by our analysis.\footnote{Note
  that this analysis assumes that the supernovae in high--density and
  low--density environments have the same luminosity distribution. If
  type Ia events are more or less luminous in lower--density regions,
  then the inferred value of $R$ would decrease or increase,
  accordingly.}

\begin{deluxetable*}{l c c l c c}
\tablewidth{0pt}
\tablecolumns{3}
\tablecaption{\label{envdist_tab} Summary of Mean Overdensities} 
\tablehead{ Sample & $<\log_{10}(1+\delta_5)>$ &
  $\sigma_{<\log_{10}(1+\delta_5)>}$ & Sample & $<\log_{10}(1+\delta_5)>$ &
  $\sigma_{<\log_{10}(1+\delta_5)>}$}
\startdata
 & & & SDSS--A & 0.070 & 0.005 \\
SN--A & -0.132 & 0.059 & SDSS--A [blue] & -0.073 & 0.007 \\
 & & & SDSS--A [red] & 0.160 & 0.006 \\ 
\hline 
SN--B & -0.023 & 0.098 & SDSS--B & 0.079 & 0.007 \\
SN--B [blue] & -0.345 & 0.053 & SDSS--B [blue] & -0.083 & 0.009 \\
SN--B [red] & 0.235 & 0.153 & SDSS--B [red] & 0.196 & 0.009 \\
\hline 
SN--C & 0.011 & 0.094 & SDSS--C & 0.100 & 0.009 \\
SN--C [blue] & -0.338 & 0.064 & SDSS--C [blue] & -0.077 & 0.012 \\
SN--C [red] & 0.248 & 0.138 & SDSS--C [red] & 0.219 & 0.012 \\
\hline 
SN--D & 0.009 & 0.101 & SDSS--D & 0.082 & 0.007 \\
SN--D [blue] & -0.171 & 0.130 & SDSS--D [blue] & -0.090 & 0.009 \\
SN--D [red] & 0.174 & 0.146 & SDSS--D [red] & 0.229 & 0.010 \\
\hline 
SN--E & -0.017 & 0.096 & SDSS--E & 0.087 & 0.007 \\
SN--E [blue] & -0.317 & 0.058 & SDSS--E [blue] & -0.074 & 0.009 \\
SN--E [red] & 0.235 & 0.153 & SDSS--E [red] & 0.216 & 0.010 \\
\hline
 & & & SDSS--F & 0.077 & 0.007 \\
 & & & SDSS--F [blue] & -0.062 & 0.009 \\
 & & & SDSS--F [red] & 0.170 & 0.009 \\
\hline 
\enddata
\tablecomments{The mean and the error on the mean of the overdensity
  distributions for the various SNe and galaxy samples.}
\end{deluxetable*}

\begin{deluxetable}{l c c}
\tablewidth{0pt}
\tablecolumns{3}
\tablecaption{\label{envdist2_tab} Summary of Statistical Tests} 
\tablehead{ Samples & $P_{{\rm WMW}}$ & $P_{{\rm KS}}$ }
\startdata
SN--A/SDSS--A & $5.5 \times 10^{-6}$ & $6.8 \times 10^{-5}$ \\
SN--A/SDSS--A [blue] & 0.011 & 0.030 \\
\hline
SN--B/SDSS--B [blue] & 0.005 & 0.006 \\
SN--B/SDSS--B [red] & 0.499 & 0.846 \\
\hline
SN--C/SDSS--C [blue] & 0.006 & 0.024 \\
SN--C/SDSS--C [red] & 0.479 & 0.632 \\
\hline
SN--D/SDSS--D [blue] & 0.068 & 0.057 \\
SN--D/SDSS--D [red] & 0.336 & 0.639 \\
\hline
SN--E/SDSS--E [blue] & 0.011 & 0.018 \\
SN--E/SDSS--E [red] & 0.479 & 0.850 \\
\hline
SN--B/SDSS--F [blue] & 0.002 & 0.005 \\
SN--B/SDSS--F [red] & 0.435 & 0.811 \\
\hline
\enddata
\tablecomments{We tabulate the $P$--values, $P_{\rm WMW}$ and $P_{\rm
    KS}$, from comparing the environment values in the listed samples,
  using the one--sided Wilcoxon--Mann--Whitney (WMW) U test and the
  two--sided Kolmogorov--Smirnov (KS) test. As discussed in \S
  \ref{sec_results}, smaller values indicate a lower probability that
  the observed differences in the samples will occur by chance if they
  are selected from the same underlying parent distribution. Note
  that, as a one--sided test, $P_{\rm WMW}$ has a maximum value of
  $0.5$.}
\end{deluxetable}

\section{Discussion}
\label{sec_disc}

In \S \ref{sec_results}, we show that the SNe Ia in blue host galaxies
occur preferentially in low--density environments relative to galaxies
of like color and luminosity (or like stellar mass and star--formation
rate). For the type Ia events in red hosts, however, we find no
significant difference between the environments of the host galaxies
and the environments of comparison galaxy samples. In the following
subsections, we examine how these results compare to the results of
related studies in the literature. We also investigate potential
selection effects, which could bias our supernova and galaxy samples,
and finally we discuss the implications of our results in terms of the
currently--unknown SN Ia progenitor population.

\subsection{Comparison to Previous Work}
\label{sec_disc_prevwork}
Recent analyses have reached a variety of conclusions regarding the
dependence of the SN Ia rate on environment. Using angular
cross--correlation techniques on data from the SNLS,
\citet{carlberg08} found that supernova hosts at $0.2 < z < 0.9$ are
more strongly clustered than a sample of field galaxies selected to
have the same redshift and ($i$--band) brightness distributions as the
hosts. However, inaccurate photometric redshifts could, by
overbroadening the measured galaxy redshift distribution, dilute the
measured angular--clustering strength for the galaxy sample relative
to that of the supernovae, for which spectroscopic redshifts were
obtained. Furthermore, the comparison sample is not selected to match
the stellar mass, rest--frame color, or SFR distributions of the host
galaxies. Thus, this result could also be attributed to the dependence
of the SN Ia rate on galaxy properties, where the SN Ia rate increases
with stellar mass and star--formation rate, and to the dependence of
galaxy properties on environment, where more massive galaxies favor
overdense environs.\footnote{There is also evidence that the most
  strongly star--forming galaxies tend to reside in dense environments
  at $z \sim 1$ \citep{cooper08a}, which might play a role in biasing
  the \citet{carlberg08} sample, since it is strongly weighted towards
  SNe at $z > 0.5$.} In fact, when weighting their field galaxy sample
by stellar mass and star--formation rate, \citet{carlberg08} show that
the clustering of the type Ia supernova hosts and field galaxies in
the SNLS are in good agreement. A parallel analysis of data drawn from
the SNLS by \citet{graham08}, focusing on SNe Ia identified within
galaxy clusters at intermediate redshift, found no significant
difference between the SN Ia rate in clusters from that in field
ellipticals. However, their supernova sample is quite small (only
three probable cluster Ia events) and the results are dominated by
statistical uncertainties.

Studying supernovae in the local $(z < 0.2)$ Universe,
\citet{sharon07} found the type Ia rate in nearby clusters to be
roughly consistent with estimates of the SN Ia rate in local
elliptical galaxies; this work, however, was based on a sample of only
six cluster supernovae and thus yielded very large uncertainties in
the measured cluster type Ia rate. A more recent analysis of data in
nearby ($z < 0.04$) clusters by \citet{mannucci08} arrived at a very
different result, finding that the SN Ia rate (per unit mass) is more
than 3 times higher in cluster ellipticals relative to field
ellipticals, using a sample of 11 cluster and 5 field type Ia events;
an identical SN rate in both samples is excluded with $P = 0.02$. In
their analysis, \citet{mannucci08} compare the rate within galaxies of
like mass and morphology, attempting to remove the known correlations
between galaxy properties and environment. For this reason, and the
similarity in redshift range probed, the \citet{mannucci08} study
provides the most significant and relevant comparison to our results.

In particular, the environment--dependent rates of \citet{mannucci08}
are most closely connected to our results regarding the environments
of red SN Ia hosts. In contrast to the results of \citet{mannucci08},
we find no significant trend such that supernovae occur more often in
overdense regions; within our SDSS samples, the environment
distribution for the subset of red hosts is indistinguishable (within
the uncertainties) from that of the samples of red comparison galaxies
(see Table \ref{envdist_tab} and Table \ref{envdist2_tab}). However,
our results regarding the environments of red host galaxies have large
statistical uncertainties associated with them, making any strong
statement about the environment dependence of the type Ia rate in red
(or elliptical) hosts impossible.

To investigate any potential discrepancy between our results and those
of \citet{mannucci08}, we attempt to test the likelihood that we would
have detected the \citet{mannucci08} result of a $> \! 3 \times$
higher SN Ia rate in cluster ellipticals relative to field
ellipticals, given our sample size of 25 red SN Ia hosts in
SN--B. From the SDSS--B galaxy sample, we select the $4256$ random red
galaxies which match the rest--frame $g-r$ color, absolute $r$--band
magnitude, and redshift distributions of the red SN Ia hosts in sample
SN--B. From this parent population, we then draw $5000$ independent
samples of $25$ random galaxies with the likelihood of drawing an
object in a high--density $(\log_{10}(1+\delta_5) > 1)$ environ forced
to be $3 \times$ greater than that of a galaxy in a low--density
$(\log_{10}(1+\delta_5) \le 1)$ environ. For each realization, we
compare the distribution of environments for the $25$ mock SN hosts to
the full sample of red galaxies in the SDSS--B sample, using the
Wilcoxon--Mann--Whitney U and Kolmogorov--Smirnov statistical
tests. Finally, this mock sample--selection analysis is repeated with
galaxies in high--density environments selected at $1 \times$, $2
\times$, and $4 \times$ the rate of galaxies in low--density regions.

\begin{figure}[h]
\centering
\plotone{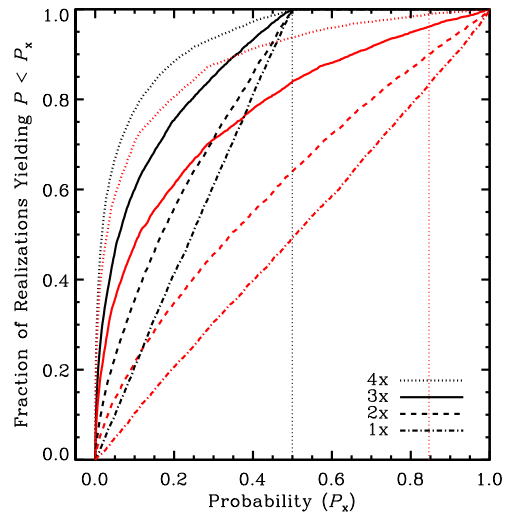}
\caption{For $5000$ mock realizations of a SN Ia rate $3 \times$
  higher in high--density than in low--density environments, we plot
  the percentage of realizations that yield a one--sided
  Wilcoxon--Mann--Whitney (WMW) probability less than $P_x$ (solid
  black line) and a two--sided Kolmogorov--Smirnov (KS) probability
  less than $P_x$ (solid red line). The black dotted, dashed, and
  dashed-dotted lines correspond to mock realizations simulating a SN
  Ia rate in high-density regions $4 \times$, $2 \times$, and $1
  \times$ that in low--density environs, respectively. Note that the
  black lines reach a fraction of unity at a probability ($P_x$) of
  $0.5$ since they correspond to the \emph{one--sided} WMW test, while
  the red lines saturate at $P_{x} = 1$ since they correspond to the
  \emph{two--sided} KS test. The black and red vertical dotted lines
  give the WMW and KS probabilities, respectively, computed from a
  comparison of the environments of red galaxies in the SN--B and
  SDSS--B samples (see Table \ref{envdist2_tab}).}
\label{mock_fig}
\end{figure}

As shown in Figure \ref{mock_fig}, in only $\sim 50\%$ of the mock
realizations with a $3 \times$ higher likelihood of SNe occurring in
dense environs, would we have rejected the hypothesis that the
environments of SN hosts are on average no more overdense than that of
the SDSS--B red galaxy sample with $P < 0.05$. The likelihood that we
would have detected a difference in the typical environment becomes
even smaller when we assume a SN Ia rate that is only $2 \times$
higher in dense environs; in this case, the WMW test $P$--value is 
$< \! 0.05$ in $25\%$ of the mock realizations.

When we apply the two--sided KS test, we find similar results; for the
majority ($> \! 60\%$ and $> \! 85\%$) of mock realizations, we would
\emph{not} have detected a significant difference in the environment
distributions at a $P > 0.05$ significance level, assuming a SN Ia
rate $3 \times$ and $2 \times$ higher in overdense regions,
respectively. Thus, while our analysis of type Ia supernovae in red
SDSS hosts does not support the conclusions of \citet{mannucci08}, our
results are not in direct conflict either, given the large
uncertainties in both results. With that said, the high WMW and KS
$P$--values computed from a comparison of the environments for the red
SN--B and SDSS--B galaxies (see Table \ref{envdist2_tab}) occur only
rarely in our simulated galaxy samples\footnote{We find $P_{\rm WMW} =
  0.499$ and $P_{\rm KS} = 0.846$ in $\sim \!  0.05\%$ and $\lesssim
  \! 5\%$ of the mock realizations with a $3\times$ higher SN rate,
  respectively.} and thus favor a SN Ia rate in cluster ellipticals
more in line with that of field ellipticals, at the low end of the
possible range given by \citet{mannucci08}.

While the \citet{mannucci08} work has a sample size roughly double
that of \citet{sharon07}, the uncertainties in their measured SN Ia
rates are still quite large, such that the field and cluster rates are
only different at a $2.1\sigma$ level, thereby leaving much
uncertainty in the dependence of the type Ia rate on environment
within early--type galaxies.

\subsection{Potential Selection Effects}
While the SDSS--II Supernova Survey includes $> \! 100$ SNe Ia with
spectroscopic follow--up in the redshift range $0.05 < z < 0.15$, less
than half of the total sample is successfully matched to a host galaxy
in the SDSS DR6 spectroscopic sample. Given the large number of
unmatched supernovae, it is important to understand any potential
biases that could arise from the particular method adopted to identify
host galaxies. For example, our observation that SNe Ia in blue
galaxies are biased towards low--density environments could naturally
arise from employing an algorithm to select host galaxies that is for
some reason biased against finding hosts in overdense regions (e.g.,
groups or clusters). Given the strong correlations between environment
and galaxy properties, it is also critical to understand any aspect of
the host--identification methodology that could be biased towards
identifying hosts of a particular galaxy color, luminosity, etc.

To test the sensitivity of our results to the particularities of the
host--identification algorithm, we define two additional SN samples,
SN--D and SN--E (see Table \ref{data_sample_tab} for sample
definitions). The SN--D sample is selected to test for any bias in the
SN--B sample towards SNe being preferentially matched to red host
galaxies. For the vast majority of SNe in the SN--B sample, only one
possible host galaxy is found within the cylindrical search window
(see \S \ref{sec_data_hosts} for the details of the SN--B matching
algorithm). However, for three SNe, multiple host galaxies are
identified within the SDSS spectroscopic data set. In Figure
\ref{cmd_fig}, we mark the location of these possible, alternate
(``secondary'') host galaxies using magenta stars, with dotted lines
connecting them to the location of the ``primary'' host, as defined in
the SN--B sample.

While the choice of the host galaxy in these ambiguous cases does not
significantly impact the measured environment of the SNe, the location
of the host in color--magnitude space directly affects whether the
host is classified as ``blue'' or ``red'' in our color cuts and
affects the composition of the comparison galaxy sample (SDSS--B).
Confusion among hosts is likely to be more common in overdense
regions,\footnote{Two of the three SNe in SN--B with an ambiguous host
  identification reside in overdense environments
  $(\log_{10}(1+\delta_5) \sim 1.5$), and the third SN resides in an
  environment that is still more overdense $(\log_{10}(1+\delta_5)
  \sim 0.36)$ than any of the SNe in the blue SN--B subsample.} which
could cause a small bias in the measured distribution of supernova
environments when divided into blue and red subsamples.

To test the robustness of our results to the ambiguity in the host
identification, we apply a Murphy's Law approach, where if anything
can go wrong to bias our SN--B sample, then it will. In the ambiguous
cases, we correspondingly match the SNe to the bluest host according
to rest--frame $g-r$ color. Given the potential host galaxies, this
creates the sample of SN hosts (SN--D) with the highest possible mean
density among blue host galaxies (see Table \ref{envdist_tab}). Even
in this extreme (and somewhat unlikely) scenario, we still find that
the distribution of environments for the blue SN hosts in sample SN--D
is distinct from that of a corresponding sample of galaxies chosen to
match in color, luminosity, and redshift (sample SDSS--D), with the
difference significant at a $> \! 94\%$, following a two--sided KS
test.

Another aspect of the host--identification methodology that might lead
to a bias against identifying hosts for SNe in dense regions is the
size of the cylindrical aperture used to search for possible host
galaxies. If this aperture is too small, then the sample could be
biased against including SNe in galaxies of greater physical size.
Massive galaxies, which are inclined to reside in overdense
environments \citep[e.g.,][]{cooper08b}, also tend to have larger
sizes \citep{shen03} and larger velocity dispersions \citep{faber76,
  dd87}, which potentially could lead to SNe occurring at projected
and velocity separations outside of the windows used to define the
hosts in SN--B.

Recognizing these correlations between galaxy size, velocity
dispersion, and environment, we alternatively identify a host galaxy
sample (SN--E) using a projected, radial window of $40\: h^{-1}$ kpc
(physical) to identify potential hosts on the plane of the sky in
conjunction with a velocity window of $\Delta v = 5000\ {\rm km}\ {\rm
  s}^{-1}$ along the line of sight, both $\sim \! 60\%$ larger than
the windows of $25\: h^{-1}$ kpc and $\Delta v = 3000\ {\rm km}\ {\rm
  s}^{-1}$ used to define the hosts in sample SN--B. Like when
defining sample SN--B, when multiple potential hosts are identified
within this window, the galaxy closest in projected distance is taken
as the host. Using these larger radial and line--of--sight search
windows, only one additional supernovae is matched to a host (relative
to the number of hosts identified in SN--B); however, the percentage
of SNe unambiguously matched to a host declines slightly from 92\% to
88\%. Thus, we conclude that our SN--B sample is not significantly
biased against SNe occurring in more massive galaxies.

A final potential selection effect associated with the identification
of SN hosts is the possibility that matching to the SDSS
spectroscopic galaxy sample is biased against matching SNe to hosts in
rich environments due to fiber collisions. Our SN--A and SN--C samples
are specifically designed to test for such an effect. By matching
directly to the imaging catalog, SN--C avoids any bias associated with
the allocation of fibers in the spectroscopic component of the SDSS.
We find that the general results obtained by analyzing the SN--B
sample are supported by the analysis of the SN--C sample. Thus, we
conclude that our results are likely robust to any bias associated
with allocation of SDSS fibers in overdense regions. As shown by
\citet{cooper05} and \citet{gerke05}, dense regions on the sky do not
always translate into dense regions in redshift space.

Looking beyond selection effects that might bias the
host--identification process, we also examine the possibility that our
supernova samples are biased against bulge--dominated systems. Given
the higher surface brightness of bulges, relative to galactic disks,
it can be more difficult to obtain spectroscopic follow--up data for a
supernova near the bulge of a galaxy. For galaxies with larger bulges,
this problem is obviously greater. Moreover, observations of local
type Ia SNe find correlations between the light--curve decline rate
and the peak luminosity with morphology, such that late--type galaxies
host brighter SNe Ia \citep[e.g.,][]{hamuy95, hamuy96,
  gallagher05}. Thus, given the correlation between morphology and
environment, where the bulge--dominated fraction increases with local
density, our observed deficit of supernovae in high--density
environments within the star--forming population and lack of an
increase in the type Ia rate within cluster red--sequence members
could both be attributed to such a morphology bias.

To test for this selection effect, we compare the morphology
distribution of our host galaxies in the SN--B sample to that of the
galaxies in the SDSS--B comparison sample. As a tracer of morphology,
we utilize the S\'{e}rsic indices as measured for SDSS galaxies by
\citet{blanton03b, blanton05a}. While the S\'{e}rsic index is a
measure of morphology derived from the fit of only a single component
to the galaxy's radial profile (e.g., versus bulge-disc
decomposition), we find no significant difference in the morphologies
of our supernova hosts relative to the morphologies of the comparison
galaxies.

In much the same way that it can be difficult to detect a supernova
superimposed on a bulge relative to a disk, follow--up spectroscopy of
candidate events (used to confirm the supernova type) is also affected
by the relative brightness of the host galaxy; spectroscopic
observations of supernovae in brighter hosts are more difficult due to
contamination of the supernova spectrum by emission from the host
galaxy. Any incompleteness in our sample that is dependent on apparent
magnitude, however, is effectively controlled for in our samples by
matching the comparison galaxy samples (e.g., SDSS--B) to the
supernova host samples (e.g., SN--B) according to color, luminosity,
and redshift. By matching both luminosity and redshift, the samples
being compared also have the same brightness. For the very same
reason, our analysis is insensitive to any incompleteness in the SDSS
spectroscopic catalog that may result from failing to obtain redshifts
for fainter galaxies.

Finally, we test the robustness of our results to the particularities
of our adopted color division between blue and red populations. To do
this, we shift the color cut given in Equation \ref{eqn1} by $\pm
0.04$ magnitudes in rest--frame $g-r$ color, which results in changes
in the blue and red components of the SN--B supernova sample of $\pm
3$ supernovae. With these changes in the subsample definitions, our
results regarding the environments of blue host galaxies remain
``highly significant'' (i.e., $P < 0.01$). Furthermore, even when
shifting the color cut by as much as $0.1$ magnitudes, the WMW U and
KS tests indicate that the blue supernova host galaxies populate a
significantly (i.e., $P \lesssim 0.05$) distinct distribution of
environments from the blue comparison galaxies in the SDSS--B sample.

\subsection{Supernova Ia Progenitors}

As discussed in \S \ref{sec_intro}, observational studies of nearby
and distant supernovae have supported the definition of two components
to the type Ia supernova rate, a ``prompt'' and a ``delayed''
component. However, there is no current observational evidence that
indicates two distinct progenitor channels associated with the two
components of the type Ia rate. With that said, recent theoretical
analyses have found difficulty reconciling observations with models in
which both the ``prompt'' and ``delayed'' Ia components result from
single--degenerate events, where a carbon--oxygen WD accretes matter
from a non--degenerate, companion star \citep[e.g.,][]{yungelson00,
  greggio05, pritchet08}. Allowing for the possibility of
double--degenerate scenarios, in which two WDs are drawn together via
angular momentum losses resulting from gravitational radiation
\citep{iben84, webbink84}, some success has been found at predicting
observed SN rates as well as the chemical enrichment of local galaxies
\citep{greggio05, matteucci06, greggio08}.

If the two components of the type Ia rate are somehow comprised of
single-- \emph{and} double--degenerate events, then the ``prompt''
component is likely primarily driven by single--degenerate events, due
to their shorter minimum timescale for occurrence. That is,
single--degenerate events are favored for the ``prompt'' component,
since the minimum timescale between formation of the progenitor star
and occurrence of the supernova is on the order of the lifetime of a
$8 {\rm M}_{\sun}$ star (i.e., $\sim \! 30$ Myr), while the minimum
timescale for occurrence of a double--degenerate event is considerably
longer \citep[$\sim \! 1$ Gyr,][]{greggio05}.  Thus, our results
suggest that single--degenerate events are favored (or are more
luminous) in low--density environs, given the assumptions detailed
above.

Now, while physical mechanisms such as galaxy mergers or harassment
are more common in high--density regions such as groups and clusters,
it is difficult to suggest a physical mechanism specific to
low--density environs that would directly impact the evolution of
stellar populations and influence the SN Ia rate or luminosity (e.g.,
by raising the binary fraction so as to produce more type Ia
events). It is seemingly more likely that environment is strongly
correlated with a galaxy property that traces differences in stellar
populations, such as stellar or gas--phase metallicity or stellar
age. In the following section (\S \ref{sec_disc_metal}), we discuss
this issue in more detail.

\subsection{The Role of Metallicity}
\label{sec_disc_metal}

As discussed in detail in \S \ref{sec_results}, we find that blue SN
Ia host galaxies are only seen in comparatively low--density regions,
suggesting that prompt supernovae Ia occur (or are found)
preferentially in underdense environs or that they are somehow
suppressed in the high--density regime. However, there is little
physical motivation for connecting large--scale environment--specific
processes (e.g., mergers, strangulation, etc.) with the generation or
suppression of type Ia supernovae. A potentially interesting, though
currently poorly unconstrained, possibility could be that
merger--induced star formation has somewhat different properties
(e.g., a different initial mass function) than typical star formation
(e.g., due to higher gas densities). The prevalence of mergers in
environments such as galaxy groups could thus be associated with an
environment dependence to the prompt Ia rate or luminosity.

Another possibility could be the suppression of SNe \emph{within}
galaxies in clusters due to ram--pressure stripping, where the
(metal--poor) outskirts of systems are preferentially confiscated by
the IGM. The stripped gas and stars would then contribute to the
intracluster light and result in intracluster (i.e., intergalactic)
supernovae \citep{galyam03, maoz05}. Such supernovae could be missed
by narrow--field supernova searches that target individual galaxies in
clusters rather than large fields \citep[e.g.,][]{cappellaro99}. The
imaging data from the SDSS--II Supernova Survey, however, covers a
large and nearly continuous field, such that any intracluster SNe
would be included in the supernova sample. Still, an intracluster SN
would be less likely to be matched to a host galaxy, thereby mimicking
a suppression of SNe within galaxies in overdense regions. Given the
large uncertainties in the cluster SN Ia rate and in the relative
contribution from intracluster events \citep{galyam03}, the role of
such supernovae remains largely unknown. With that said, the
relatively low concentration of blue galaxies in local clusters helps
to minimize the impact of intracluster events on our results, as
related to the preferential occurrence of SNe Ia in blue host galaxies
in underdense environs.

Alternatively, it could be the case that environment is correlated
with a galaxy property or properties, for which we did not control in
our analysis. A correlation between a given physical property and
environment would cause the overdensities about our SN host galaxies
to be skewed relative to the comparison sample, if supernova host
galaxies are a (relatively) biased tracer of that galaxy property. One
likely galaxy characteristic to consider is metallicity. 

The metal abundance of a galaxy is often quantified in two ways: [1]
by measuring the metal abundance in the interstellar medium (the
gas--phase metallicity) and [2] by assessing the amount of metals
locked up in the stellar population (the stellar metallicity). The
gas--phase metallicity is a product of the recent ($< \! 1$ Gyr)
accretion and star--formation history of the galaxy
\citep{finlator08}. In contrast, the stellar metallicity, which is
commonly inferred from fits to stellar absorption features in optical
spectra, traces the metals locked up in the old stellar
population. Thus, stellar metallicity is a tracer of the integrated
star--formation history and a measure of the gas--phase metallicity in
the galaxy when the bulk of the stars formed. Here, we investigate the
potential role of both stellar and gas--phase metallicity in biasing
our supernova samples in star--forming galaxies towards underdense
environments.

To study the potential role of stellar metallicity in our results, we
employ the measurements of \citet{gallazzi05}, which are based on
model fits to spectral absorption features in the SDSS DR4 spectra.
The measurements are sensitive to the signal--to--noise ratio of the
spectrum \citep[see Table 1 of][]{gallazzi05}, which limits our
ability to directly constrain the stellar metallicities of the SN
hosts. More than half of the SNe in our sample are at $z > 0.1$, which
means that the host galaxies tend to be relatively faint in the
$r$--band (less than $25\%$ of the SNe have an $r$--band magnitude $<
\! 16.5$) and the derived metallicity values are highly uncertain. For
this reason, we are unable to make any meaningful statements about the
stellar metallicity values of the hosts relative to the comparison
galaxy samples.

However, we are able to study the relationship between stellar
metallicity, $Z$, and environment for the general SDSS galaxy
population, which allows us to understand how our results regarding
the environments of type Ia hosts could be understood in terms of a
stellar metallicity bias. In the Appendix, we investigate in detail
the relationship between metallicity and environment for three
magnitude--limited samples drawn from the \citet{gallazzi05} catalog.
As shown in Figure \ref{zenv2_fig}, we find a significant correlation
such that galaxies with more metal--rich stellar populations typically
reside in more overdense environs. Furthermore, this general
correlation persists when focusing on just those galaxies that reside
on the blue cloud (following Equation \ref{eqn1}); although, the trend
is considerably weaker.

A corresponding relationship between gas--phase metallicity and
environment was recently published by \citet{cooper08b}; studying
star--forming galaxies in the SDSS DR4, they found a significant
correlation between average gas--phase metallicity and local galaxy
density, such that more metal--rich galaxies favor regions of higher
overdensity. Along the blue cloud, this metallicity--density relation
is comparable in strength to the well--known color--density
relation. Moreover, \citet{cooper08b} show that metallicity has a
relationship with environment separate from that observed with color
and luminosity (or with stellar mass). Gas--phase metallicity is
somewhat unique in this regard, as other galaxy properties (e.g.,
surface brightness, S\'{e}rsic index, or stellar mass) that are
strongly correlated with overdensity do not show a relationship with
environment separate from that observed with color and luminosity
\citep{blanton05a, cooper08b}.

In order for the results in \S \ref{sec_results} to be due to either
of these observed correlations between environment and metallicity
(either stellar or gas--phase), the blue SN host galaxies must be
biased towards lower stellar or gas--phase metallicities than the
comparison galaxy samples. As discussed above, however, we are unable
to directly test for any potential bias in our supernova samples, due
to the lack of metallicity information (both stellar and gas--phase)
for the majority of the hosts. Given the low signal--to--noise of the
SDSS spectra, reliable metallicity measures are not feasible. Without
direct constraints on the metallicities of the SDSS host galaxies, our
results regarding the environments of SNe Ia in star--forming galaxies
could still (at least partially) be understood in terms of a
metallicity effect if the luminosity or the rate of type Ia events
depends on metallicity, such that intrinsically brighter supernovae
arise from more metal--poor progenitors or such that the type Ia rate
is elevated in metal--poor host galaxies.

However, recent observational work by \citet{howell08}, studying SNe
Ia in a relatively large sample $(> \!  100)$ of star--forming and
quiescent host galaxies at $z \le 0.75$, suggests that variation in
gas--phase metallicity only accounts for a small portion $(< \! 10\%)$
of the measured dispersion in type Ia luminosities \citep[see Figures
5, 7, and 11 of][]{howell08}. A variety of previous studies
\citep[e.g.,][]{hamuy00, ivanov00, gallagher05}, using both gas--phase
and stellar metallicity estimates, also found no significant evidence
for a correlation between metallicity and the properties of SNe Ia.
With that said, whether low--metallicity stars might yield more
luminous type Ia events is still a relatively poorly constrained
question. For example, the gas--phase metallicity estimates employed
by \citet{howell08} are derived from using measurements of the hosts'
stellar masses to estimate the oxygen abundances according to the
median mass--metallicity relation for nearby star--forming systems
\citep{tremonti04}. This method is applied even to those systems
thought to be quiescent and are thus not included in the analysis of
\citet{tremonti04}. Future observations, yielding direct oxygen
abundance measurements, are needed to better determine the
relationship between metallicity and type Ia luminosity.

While the luminosities of type Ia SNe show no significant correlation
with metallicity, our observations of supernova environments could
also be explained by an increase in the SN Ia rate in metal--poor
hosts. From an observational standpoint, the dependence of the type Ia
rate on host metallicity remains unconstrained, with current supernova
samples including precision metallicity measurements for only a small
portion of the host population. Still, recent observational studies
show that a considerable number of type Ia supernovae have been found
in low--metallicity galaxies \citep{strolger02, prieto08}. On the
other hand, observations of type Ia supernovae in nearby,
star--forming galaxies with significantly enriched interstellar media
are evidence that the SN Ia rate is not zero in the metal--rich regime
\citep[e.g.,][]{gallagher05}.

In addition, theoretical models of supernovae suggest that the type Ia
rate depends significantly on metallicity, such that the rate is lower
in galaxies with lower metallicities \citep[e.g.,][but also see
\citealt{umeda99a}]{tornambe86, hachisu96, kobayashi98,
  kobayashi00}. This theoretical metallicity effect, however, works
counter to that which would explain our observations of supernova and
galaxy environments, making our results regarding the environments of
type Ia SNe in star--forming systems even more remarkable. With that
said, the theoretical models are generally constrained on a limited
basis, being forced to match observations of chemical evolution in the
solar neighborhood.

In spite of the theoretical predictions regarding the dependence of
the SN Ia rate on metallicity, our results appear to be most easily
understood in terms of a gas--phase metallicity effect, where prompt
SNe Ia preferentially arise from metal--poor progenitors. In
particular, the sharp cut--off in the observed distribution of
environments for SNe Ia in blue hosts could be the result of a sharp
feature in the metallicity distribution of the prompt SNe Ia
progenitor population, such that prompt type Ia events rarely result
from the evolution of relatively metal--rich stars.

This picture is supported (though circumstantially) by several key
points. First, the bias in the galaxy environment distribution for SNe
Ia in star--forming systems towards underdense regions is likely
attributable to the prompt (versus delayed) component of the type Ia
population, since delayed SNe are seen in both star--forming and
quiescent galaxies and we find no evidence for any dependence of the
type Ia rate on environment within red, non--star--forming
galaxies. Furthermore, weak evidence exists \citep{mannucci08} to
suggest that the SN Ia rate is actually higher among elliptical
galaxies in clusters versus the field, which would work counter to the
trend we observe in star--forming hosts, thereby suggesting that
delayed and prompt SNe Ia have opposing relationships with galaxy
environment, such that prompt events, which dominate in the
star--forming galaxy population, favor low--density environs and
delayed events, which comprise all of the SNe observed in quiescent
systems, are preferentially found in high--density environs.

There are two significant reasons that gas--phase metallicity (and not
another galaxy property such as stellar metallicity or age) is likely
driving our observational results. As discussed above, the prompt SN
Ia component is correlated with star formation and thus thought to
result from the evolution of more massive stars (perhaps with masses
of $\gtrsim 8 {\rm M}_{\sun}$). For this reason, the metallicity of
the progenitor population is more likely to be connected to the
gas--phase metallicity and not stellar metallicity. 

In addition, in our analysis, we compare the environments of our SN
host galaxies to samples of galaxies with matched rest--frame color,
luminosity, stellar mass, and star--formation rate distributions. {\bf
  Thus, for our results to be connected to a particular physical
  property, then that property must have a relationship with
  environment separate from that observed with color and luminosity
  (or stellar mass or star--formation rate) within the star--forming
  population. Gas--phase metallicity is the only galaxy characteristic
  known to have such a relationship with environment
  \citep{cooper08b}.} As discussed in more detail in the Appendix,
stellar metallicity shows no relationship with environment separate
from that observed with color and luminosity along the blue
cloud. Moreover, even though type Ia luminosities may depend on
stellar age in some systems \citep[e.g., early--type
systems,][]{gallagher08}, we also find that luminosity--weighted mean
stellar age exhibits no significant correlation with environment at
fixed color and luminosity among the star--forming population (see
Appendix).\footnote{Morphology and surface brightness likewise show no
  relationship with galaxy density separate from that observed between
  color and luminosity \citep{blanton05a}.}

While our results appear to be the manifestation of a metallicity bias
in our supernova host samples relative to the comparison galaxy
samples, such that the hosts are more metal--poor and the type Ia rate
(or luminosity) is higher at lower metallicity, the exact physical
explanation for why the SN Ia rate (or luminosity) would be elevated
at low metallicities remains unaddressed. One possible explanation for
an increase in the SN Ia rate would be variation in the stellar
initial mass function (IMF) with metallicity. However, measurements of
the IMF in nearby star--forming regions show that it does not depend
on metallicity down to very low masses \citep[$\lesssim 0.1 {\rm
  M}_{\sun}$,][]{bate05, yasui06, yasui08}. In addition, there is no
observational evidence to suggest that the IMF varies with galaxy
environment; instead, the stellar IMF is generally thought to be
universal \citep[at least locally, e.g.,][]{elmegreen99, kroupa07,
  selman08}.

Alternatively, a decrease in the SN Ia rate with metallicity could be
attributed to metallicity--dependent variation in the binary
fraction. Within the standard model for type Ia supernovae, an
increase in the binary fraction should lead to an increase in the SN
rate. However, there is no evidence to suggest that the binary
fraction shows any variation with metallicity (or environment) across
a broad range of metallicities in the local Universe \citep[e.g.,][but
see also \citealt{machida08, machida09}]{carney05}. Furthermore, type
Ia supernovae might not result solely from the evolution of binary
systems. As discussed in more detail by \citet{maoz08} and
\citet{tout05}, our knowledge of type Ia supernova progenitors is
quite poor, and a ``single--star'' SN Ia channel could help explain
some observations of type Ia events that remain poorly understood
within the standard binary--driven model.

Finally, as highlighted throughout this paper, our results are also
consistent with an increase in the type Ia luminosity (rather than an
increase in the rate) at low metallicities. While some early
one--dimensional simulations of type Ia explosions suggested that
there is a positive (or perhaps no) correlation between metallicity
and type Ia luminosity \citep[e.g.,][]{umeda99b, dominguez01}, more
recent theoretical analysis spanning a broader range of metallicities
support a picture in which more metal--poor progenitors yield more
luminous type Ia events \citep{timmes03, townsley09, kasen09}. In
particular, the analytical work of \citet{timmes03} concludes that the
peak luminosity of type Ia events is anti--correlated with the
metallicity of the progenitor, such that the $^{56}$Ni mass, a proxy
for luminosity \citep{arnett82, pinto00}, produced in a type Ia event
decreases linearly with increasing metallicity --- i.e., luminosity
depends roughly linearly on the gas--phase metallicity in the case of
supernovae with short delay times, where the current gas--phase
metallicity reflects the stellar metallicity of the progenitor. So,
while there is currently no observational evidence for a correlation
between type Ia luminosity and gas--phase metallicity within
star--forming host galaxies \citep[i.e., among prompt SNe
Ia,][]{howell08}, there is theoretical support to suggest that our
results could be understood in terms of a luminosity effect in lieu of
a bias in the type Ia rate.

\section{Summary}

In this paper, we present an analysis of the large--scale environments
of type Ia supernovae in the local $(0.05 < z < 0.15)$ Universe, using
data drawn from the SDSS--II Supernova Survey and from the SDSS DR6
database. We estimate the local overdensity about each galaxy
according to the projected fifth--nearest--neighbor surface
density. For a range of supernova host galaxy subsamples, we define
comparison galaxy subsamples, selected to match the color, luminosity,
and redshift or stellar mass, star--formation rate, and redshift of
the hosts. Using a variety of statistical tests, we then compare the
distribution of environments for the supernova hosts and the
comparison galaxies. Our principal results are as follows.

\begin{enumerate}

\item We find that type Ia supernovae in blue (i.e., star--forming)
  host galaxies tend to reside in low--density environments relative
  to galaxies of similar stellar mass and star--formation rate (or
  similar rest--frame color and luminosity). This analysis represents
  the first constraint on the environment dependence of the SN Ia rate
  in star--forming systems.

\item We find no difference between the environment distributions of
  red host galaxies and galaxies of like stellar mass and
  star--formation rate, within the observational uncertainties. This
  result is in some contrast to recent observational work by
  \citet{mannucci08}, which found a SN Ia rate more than $3 \times$
  higher in local cluster ellipticals relative to field ellipticals,
  though within the errors of each measurement. While our sample sizes
  are still somewhat small ($\lesssim \! 30$ Ia events in red hosts),
  tests on simulated galaxy catalogs suggest that the distribution of
  environments for red, SN Ia hosts presented herein is in poor
  agreement with a cluster SN Ia rate as strongly elevated relative to
  the field rate as \citet{mannucci08} find.

\item We find a strong cut--off, at $\log_{10}(1+\delta_5) \sim 0.15$,
  in the observed distribution of environments for type Ia supernovae
  in star--forming galaxies. Statistical tests show that such a strong
  cut--off in overdensities is likely to result from a type Ia SN rate
  that is $\gtrsim \! 2\times$ higher within star--forming galaxies in
  low--density environments relative to like systems in more overdense
  regions.

\item We conclude that variation in gas--phase metallicity is the most
  likely explanation for the observed difference between the
  environments of type Ia SNe in star--forming galaxies and galaxies
  with like color, luminosity, stellar mass, and star--formation rate;
  type Ia events are more luminous or more numerous in metal--poor
  galaxies. Given existing observational constraints, which indicate a
  lack of type Ia luminosity dependence on metallicity, our analysis
  suggests that prompt type Ia events preferentially result from the
  evolution of relatively metal--poor stars (i.e., have a higher rate
  in metal--poor systems).

\item Theoretical models of exploding white dwarfs likely need to be
  revisited, since they predict a strong fall off in the type Ia rate
  at low metallicity, while our observations unmistakably suggest the
  opposite effect for the prompt SN Ia channel.

\item We predict an increase in prompt type Ia supernovae at higher
  redshift, as star--formation activity increases and metallicity
  decreases.

\end{enumerate}


\acknowledgments Support for this work was provided by NASA through
the Spitzer Space Telescope Fellowship Program. M.C.C.\ would like to
thank Michael Blanton and David Hogg for their assistance in utilizing
the NYU--VAGC data products. This work benefited greatly from
conversations with John Hillier, Tom Matheson, Dave Sand, Amelia
Stutz, Dean Townsley, Christy Tremonti, and Michael Wood--Vasey.
Finally, the authors would like to thank John Moustakas for providing
comparison data that enabled a direct check of our emission--line
measurements.

Funding for the SDSS has been provided by the Alfred P.\ Sloan Foundation,
the Participating Institutions, the National Science Foundation, the
U.S.\ Department of Energy, the National Aeronautics and Space
Administration, the Japanese Monbukagakusho, the Max Planck Society, and
the Higher Education Funding Council for England. The SDSS Web Site is
http://www.sdss.org/.

The SDSS is managed by the Astrophysical Research Consortium for the
Participating Institutions. The Participating Institutions are the American
Museum of Natural History, Astrophysical Institute Potsdam, University of
Basel, University of Cambridge, Case Western Reserve University, University
of Chicago, Drexel University, Fermilab, the Institute for Advanced Study,
the Japan Participation Group, Johns Hopkins University, the Joint
Institute for Nuclear Astrophysics, the Kavli Institute for Particle
Astrophysics and Cosmology, the Korean Scientist Group, the Chinese Academy
of Sciences (LAMOST), Los Alamos National Laboratory, the
Max--Planck--Institute for Astronomy (MPIA), the Max--Planck--Institute for
Astrophysics (MPA), New Mexico State University, Ohio State University,
University of Pittsburgh, University of Portsmouth, Princeton University,
the United States Naval Observatory, and the University of Washington.


\appendix

\begin{figure}[h]
\centering
\plotone{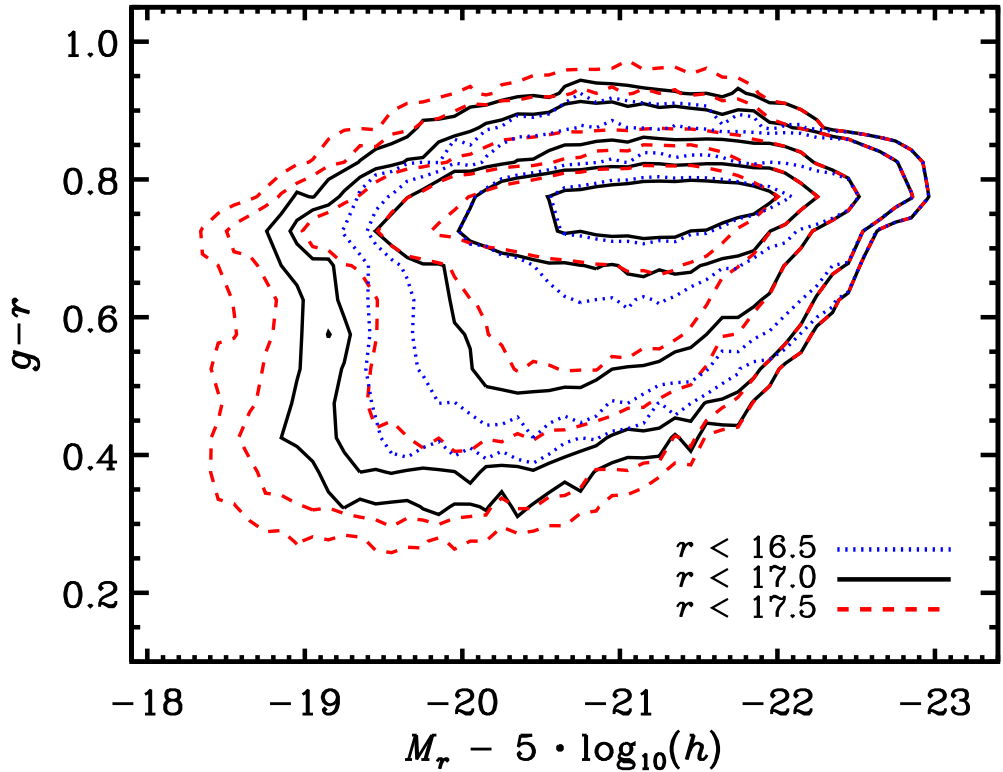}
\caption{The rest--frame $g-r$ color and $r$--band absolute magnitude
  distributions for magnitude--limited samples drawn from 
  \citet{gallazzi05}.}
\label{zcmd_fig}
\end{figure}

\begin{figure}[h]
\centering
\plotone{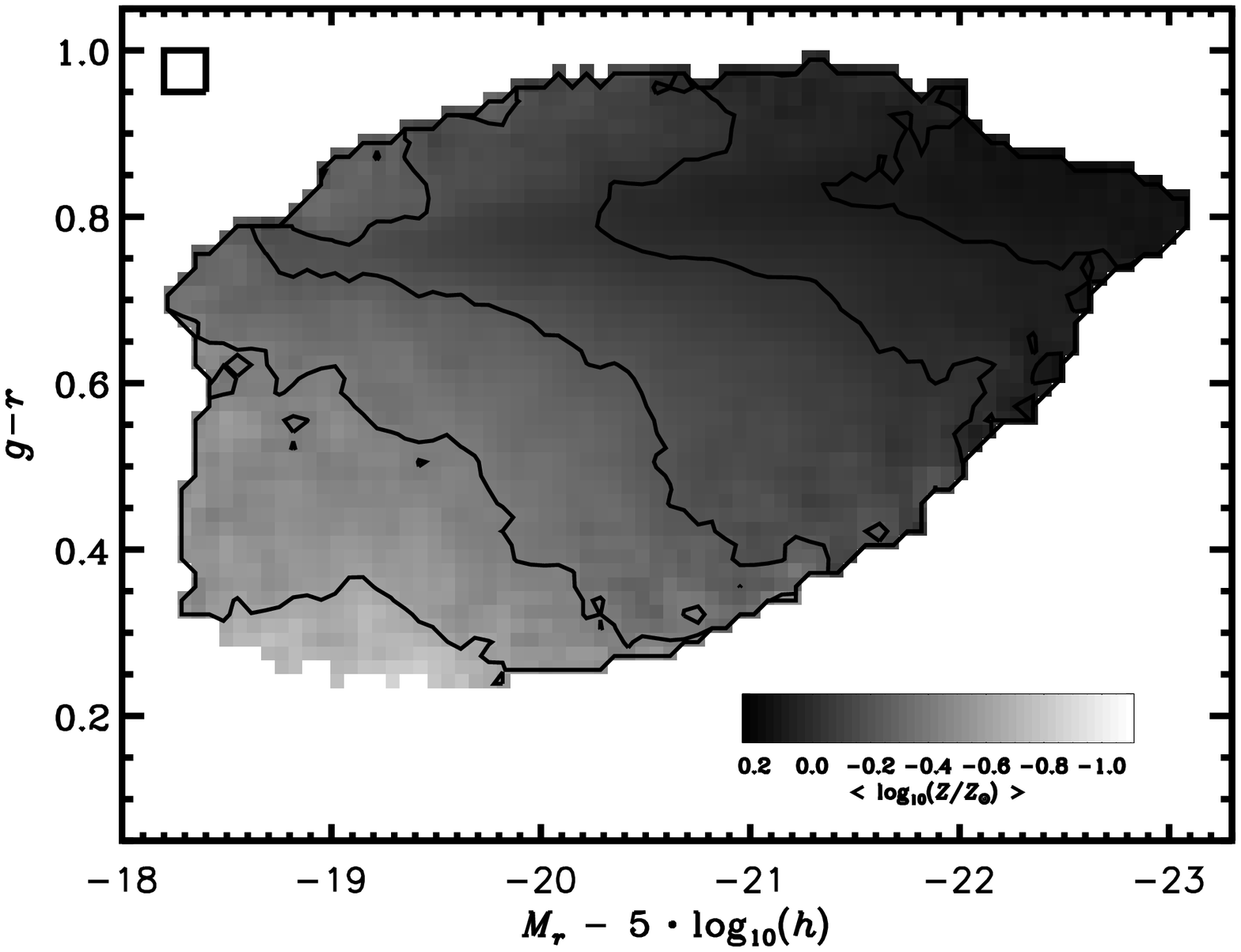}
\caption{The mean stellar metallicity, $Z$, as a function of
  rest--frame $g-r$ color and $r$--band absolute magnitude for
  galaxies brighter than $r = 17.5$ in the catalog of
  \citet{gallazzi05}. The mean metallicity is computed in a sliding
  box of width $\Delta M_r = 0.2$ and height $\Delta(g-r) = 0.05$, as
  shown in the upper left corner.}
\label{avgz_fig}
\end{figure}

To study the relationship between stellar metallicity, $Z$, and
environment for the general SDSS galaxy population, we select three
subsamples from the \citet{gallazzi05} catalog down to limiting
magnitudes of $r = 16.5, 17, 17.5$, which test for the dependence of
our results on the precision of the metallicity measurements while
also probing increasingly more intrinsically faint samples. Figure
\ref{zcmd_fig} shows the distribution of the three magnitude--limited
samples in $g-r$ versus $M_r$ color--magnitude space. Extending to
fainter samples includes more metal--poor systems, especially at the
faint, blue end of the blue cloud (see Figure \ref{avgz_fig}).

\begin{figure}[h]
\centering
\plotone{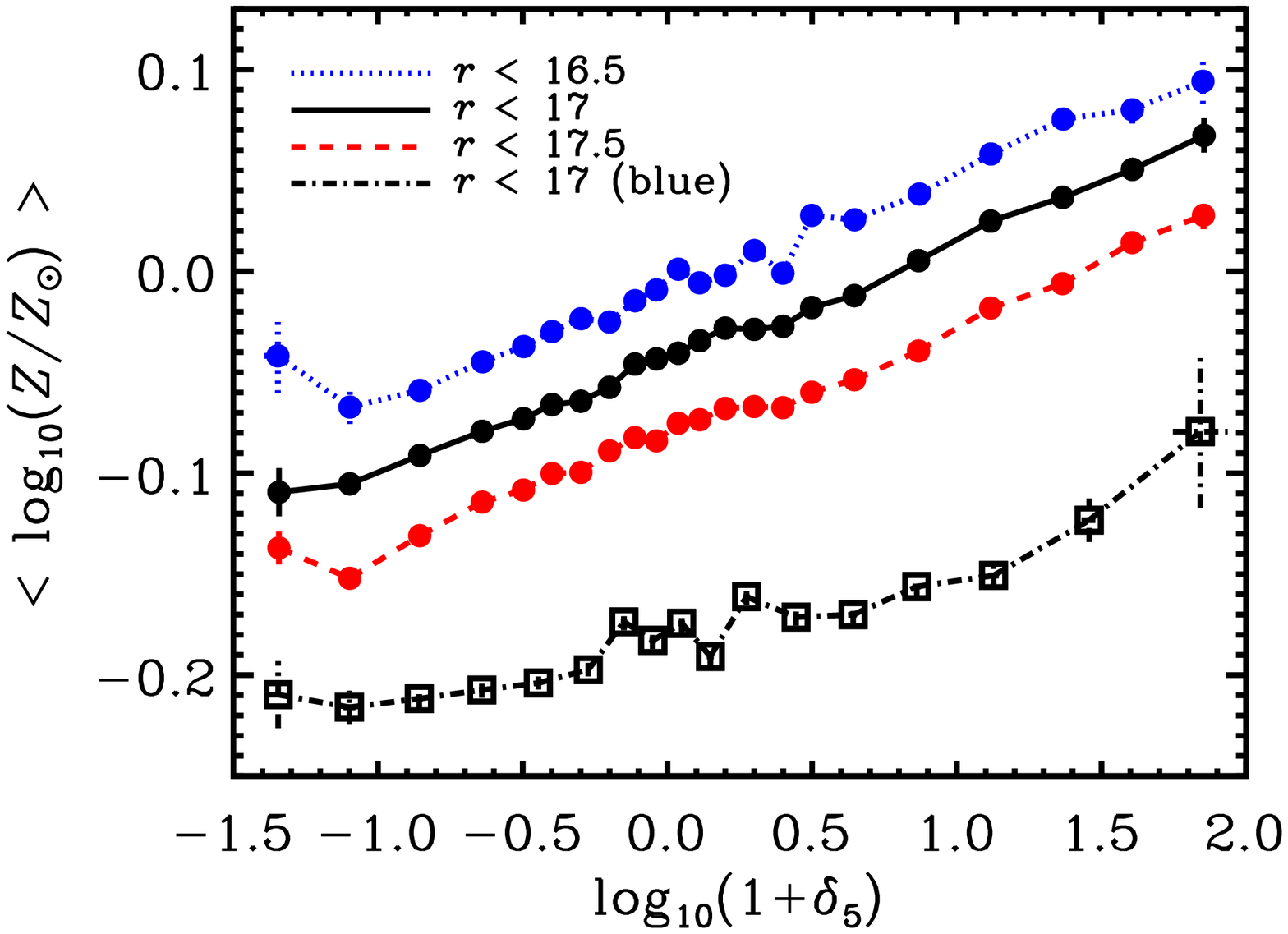}
\caption{The mean stellar metallicity, $< Z >$, as a function of local
  galaxy overdensity, $\log_{10}(1+\delta_5)$, in magnitude--limited
  samples drawn from \citet{gallazzi05}.}
\label{zenv2_fig}
\end{figure}

For these three magnitude--limited samples, we find a significant
correlation between stellar metallicity and environment such that
galaxies with more metal--rich stellar populations typically reside in
more overdense environs, with the trend relatively independent of the
magnitude limit of the sample (see Figure \ref{zenv2_fig}). This
general correlation persists when focusing on just those galaxies that
reside on the blue cloud (following Equation \ref{eqn1}); however, the
trend is considerably weaker.

As stated in \S \ref{sec_disc_metal}, for the main observational
results of this paper to be connected to a particular physical
property, that property must have a relationship with environment
separate from that observed between color and luminosity (or stellar
mass and star--formation rate). To test whether stellar metallicity
has such a relationship with environment, we remove the dependence of
environment on $g-r$ rest--frame color and $r$--band luminosity for a
magnitude--limited $(r < 17)$ sample of galaxies on the blue cloud,
using the methodology detailed by \citet{cooper08b}. In short, we
subtract the mean overdensity at the color and luminosity of each
galaxy from the measured overdensity:

\begin{equation}
  \Delta_5 = \log_{10}(1+\delta_5) - < \log_{10}(1+\delta_5)[g-r, M_r]
  > ,
\label{eqn_resid}
\end{equation}
where  the distribution of mean environment with color and absolute
$r$--band magitude, $< \log_{10}(1+\delta_5)[g-r, M_r] >$, is median
smoothed on $\Delta(g-r) = 0.15$ and $\Delta M_r = 0.6$ scales prior to
subtraction. This difference gives the ``residual'' environment,
$\Delta_5$, which quantifies the overdensity about a galaxy relative
to galaxies of like color and luminosity. As shown in
\citet{cooper08b}, this new measurement of environment shows no
correlation with color and luminosity (by design) or stellar mass and
star--formation rate. 

\begin{figure}[h]
\centering
\plotone{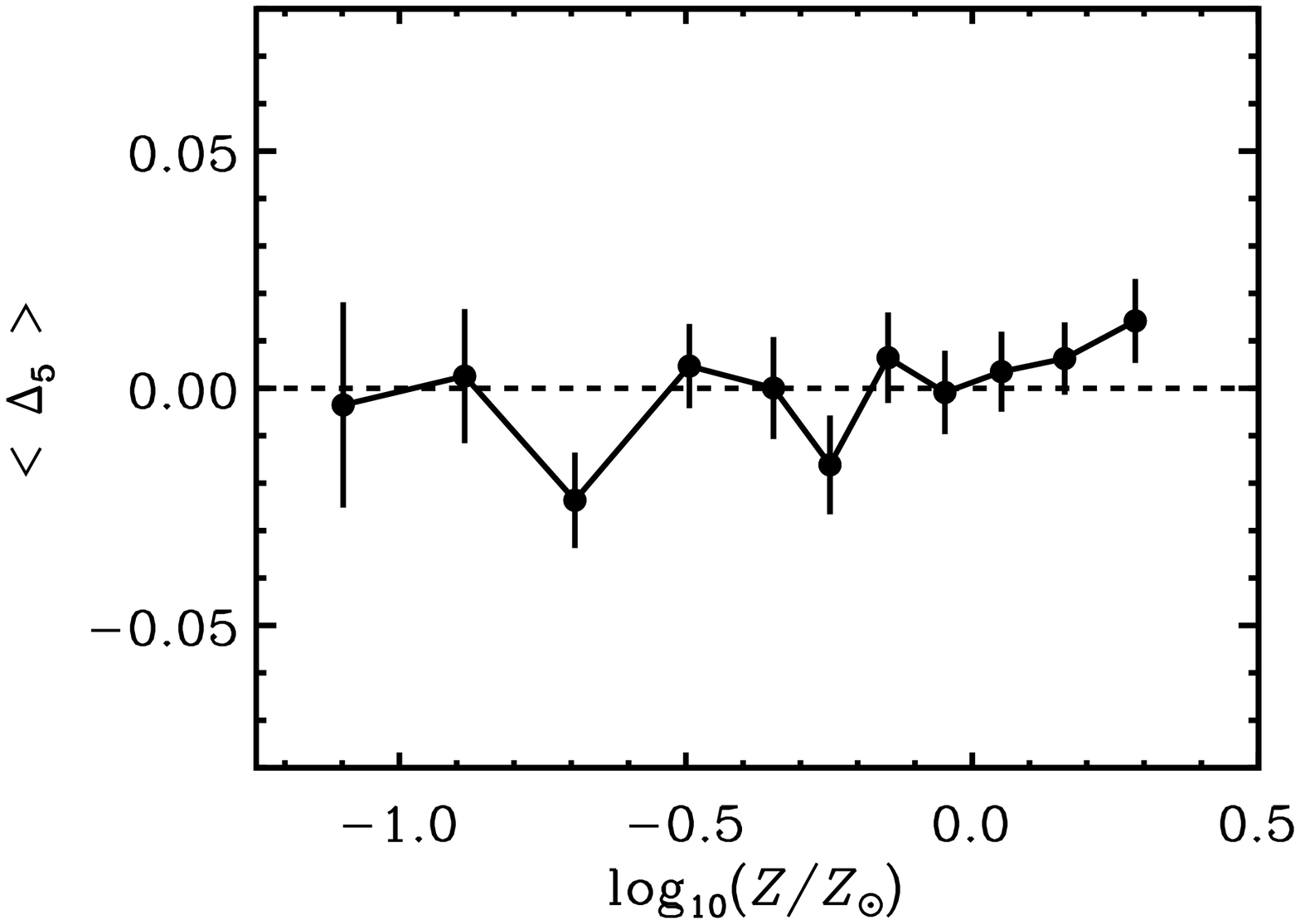}
\caption{The dependence of mean residual environment, $\Delta_5$, on
  stellar metallicity for the blue galaxies in a magnitude--limited
  $(r < 17)$ sample drawn from \citet{gallazzi05}. The residual
  environment is computed following the methods of \citet{cooper08b}.
  We find no relationship between stellar metallicity and environment
  beyond that observed between color, luminosity, and environment. }
\label{zmetal_resid}
\end{figure}

We then study the relationship between residual environment and
stellar metallicity, for our sample of star--forming galaxies with $r
< 17$. As shown in Figure \ref{zmetal_resid}, we find no significant
residual trend between stellar metallicity and environment. That is,
the mean residual environment is independent of stellar metallicity
over the full range of metallicities probed. Thus, like morphology and
surface brightness, stellar metallicity does \emph{not} have a
relationship with environment separate from that observed with color
and luminosity.

For the same sample of blue galaxies (with $r < 17$), we also examine
the correlation between luminosity--weighted mean stellar age and
environment. Again, we use the age measurements of \citet{gallazzi05},
which are based on model fits to the spectra features in the SDSS
spectra. The model spectra utilized by \citet{gallazzi05} are derived
from the population synthesis models of \citet{bc03} and span a broad
range of star--formation histories. The models are simultaneously fit
to a minimum set of metal-- and age--sensitivie spectral indices
(D4000, H$\beta$, H$\delta_A$ + H$\gamma_A$, [Mg$_{2}$Fe], and
[MgFe]$^{\prime}$), yielding measurements of age and metallicity with
typcial uncertainties on the order of $\sigma_t \lesssim 0.15$ dex and
$\sigma_Z \lesssim 0.3$.

\begin{figure}[h]
\centering
\plotone{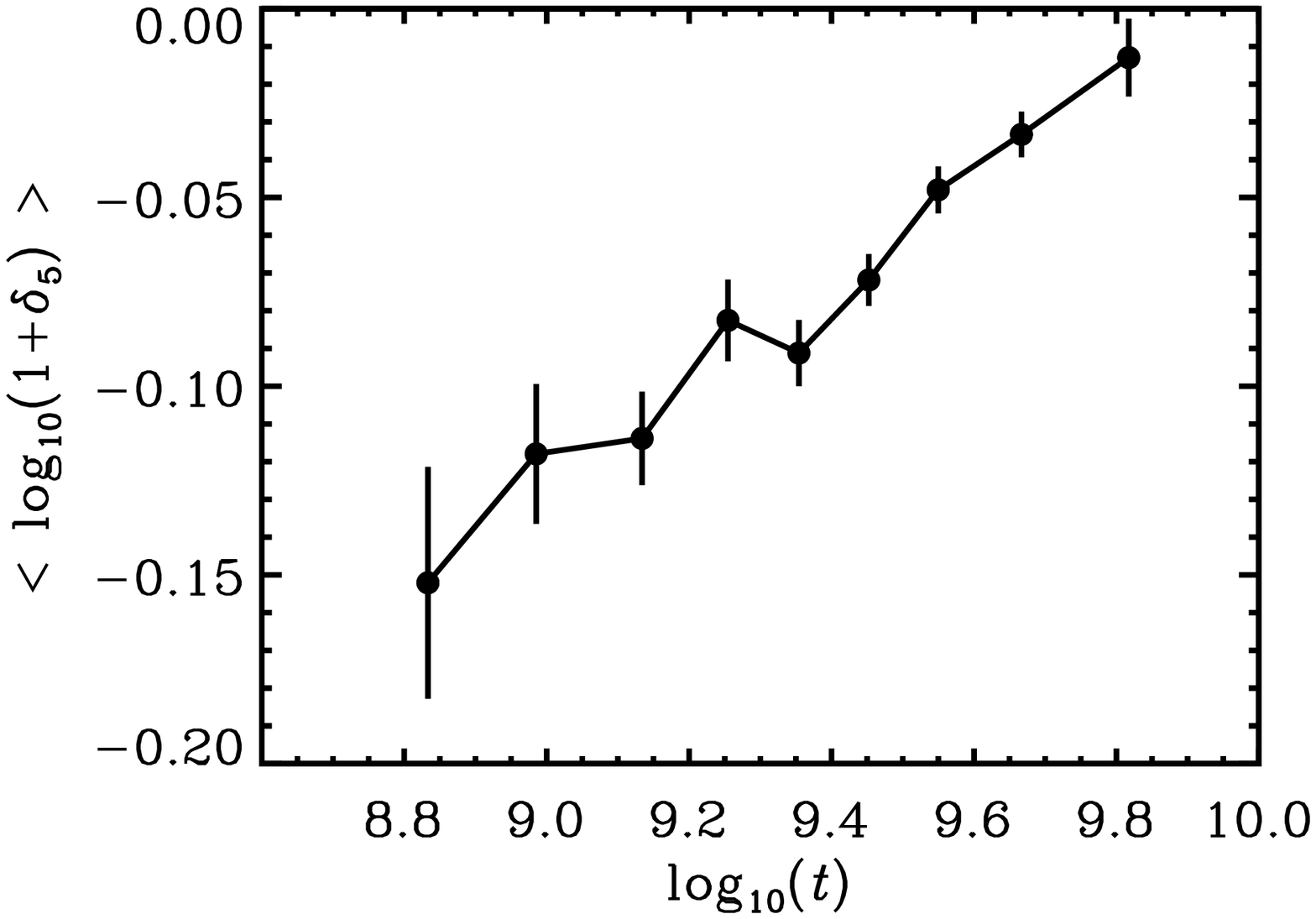}
\caption{The dependence of mean absolute environment,
  $log_{10}(1+\delta_5)$, on stellar age, $(t)$, for the blue galaxies
  in a magnitude--limited $(r < 17)$ sample drawn from
  \citet{gallazzi05}. We find a significant relationship between mean
  stell age and environment such that older (i.e., more massive)
  galaxies favor overdense regions. }
\label{age_absolut}
\end{figure}

As shown in Figure \ref{age_absolut}, we find that there is a
significant correlation between stellar age and absolute local
environment along the blue cloud; galaxies with older stellar
populations tend to reside in overdense environs relative to their
counterparts with younger stellar composition. This effect is far from
surprising given the correlation between stellar age and mass, where
more massive systems tend to have more enriched and older stellar
populations \citep[see Fig.\ 8 of ][]{gallazzi05}.

\begin{figure}[h]
\centering
\plotone{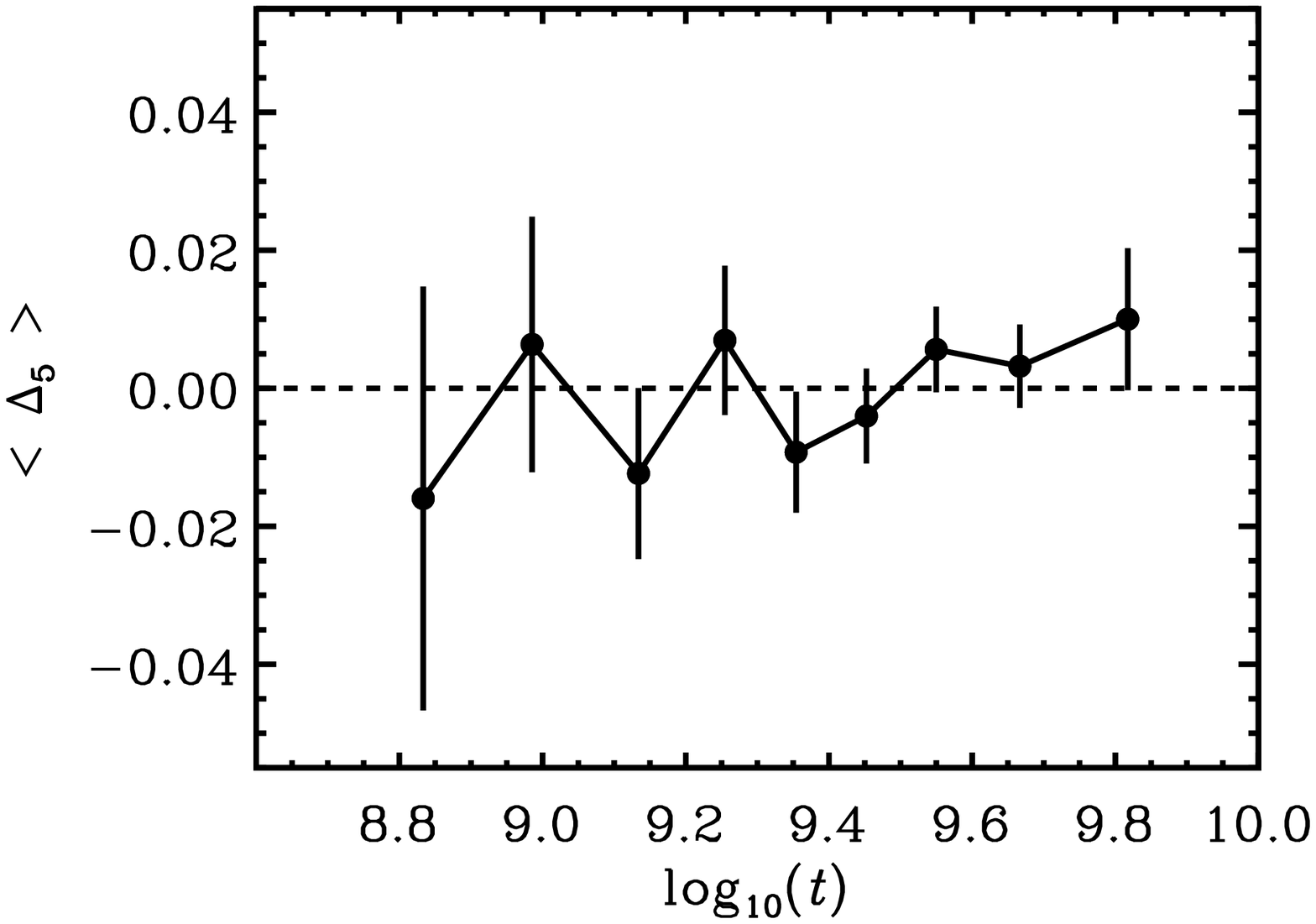}
\caption{The dependence of mean residual environment, $\Delta_5$, on
  stellar age for the blue galaxies in a magnitude--limited $(r < 17)$
  sample drawn from \citet{gallazzi05}. The residual environment is
  computed following the methods of \citet{cooper08b}. We find no
  relationship between stellar age and environment beyond that
  observed between color, luminosity, and environment. }
\label{age_resid}
\end{figure}

Now, we remove the mean dependence of environment on luminosity and
color, again following Equation \ref{eqn_resid}, and study the
correlation between residual environment, $\Delta_5$, and stellar
age. We find that there is no sigificant relationship between age and
residual environment (see Figure \ref{age_resid}). That is, stellar
age --- like stellar metallicity --- does \emph{not} have a
relationship with environment separate from that observed with color
and luminosity.

\end{document}